\documentclass[
reprint,
superscriptaddress,
frontmatterverbose, 
showpacs,
preprintnumbers,
nofootinbib,
amsmath,
amssymb,
aps,
 onecolumn,
floatfix,
]{revtex4-2}
\pdfoutput=1

\usepackage[dvipsnames]{xcolor}
\usepackage{multirow}
\usepackage{graphicx}
\usepackage{adjustbox}
\usepackage{booktabs}
\usepackage{appendix}
\usepackage{slashed}
\usepackage{url}
\usepackage{xspace}
\usepackage{braket}
\usepackage{tabularx}
\usepackage[normalem]{ulem} 
\usepackage{siunitx}
\usepackage[ISO]{diffcoeff}
\usepackage[utf8]{inputenc}
\usepackage{xspace}
\usepackage{bm}
\definecolor{darkblue}{rgb}{0,0,0.5}
\definecolor{darkgreen}{rgb}{0.0,0.5,0.2}
\definecolor{darkred}{rgb}{0.6,0,0}
\usepackage[pdfstartview=XYZ,
bookmarks=true,
colorlinks=true,
linkcolor=darkred,
urlcolor=blue,
citecolor=darkgreen,
pdftex,
bookmarks=true,
breaklinks=true,
linktocpage=true, 
hyperindex=true
]{hyperref}
\usepackage[capitalise]{cleveref}
\usepackage{orcidlink}

\usepackage{fullpage}
\usepackage{geometry}      
\AtBeginDocument{
  \heavyrulewidth=.08em
  \lightrulewidth=.05em
  \cmidrulewidth=.03em
  \belowrulesep=.65ex
  \belowbottomsep=0pt
  \aboverulesep=.4ex
  \abovetopsep=0pt
  \cmidrulesep=\doublerulesep
  \cmidrulekern=.5em
  \defaultaddspace= .5em
  \setlength{\tabcolsep}{0.5em}
}

\newcommand{\IC}{IceCube }
\newcommand{\ICns}{IceCube}

\begin{document}

\title{Blazar Boosted Dark Matter in IceCube}

\author{Alberto M. Gago}
\email{agago@pucp.edu.pe}
\affiliation{Secci\'on F\'isica, Departamento de Ciencias, Pontificia Universidad Cat\'olica del Per\'u, Apartado 1761, Lima, Per\'u}
\author{Jaime Hoefken Zink}
\email{jaime.hoefkenzink@ncbj.gov.pl}
\affiliation{National Centre for Nuclear Research, Pasteura 7, Warsaw, PL-02-093, Poland} 
\author{Joel Jones-Pérez}
\email{jones.j@pucp.edu.pe}
\affiliation{Secci\'on F\'isica, Departamento de Ciencias, Pontificia Universidad Cat\'olica del Per\'u, Apartado 1761, Lima, Per\'u} 
\author{Gabriel Damián Zapata}
\email{gabriel.zapata@correo.nucleares.unam.mx}
\affiliation{Secci\'on F\'isica, Departamento de Ciencias, Pontificia Universidad Cat\'olica del Per\'u, Apartado 1761, Lima, Per\'u} 
\affiliation{Instituto de Ciencias Nucleares, Universidad Nacional Autónoma de México, 04510 Ciudad de México, Mexico}

\date{\today}

\begin{abstract}
We study the sensitivity of \IC to blazar-boosted dark matter in a fermionic dark matter model with a massive vector mediator coupling to quarks. To this aim, we compute the diffuse flux arising from a sample of 324 blazars with proton spectra inferred from multiwavelength observations, adopting conservative dark matter spike profiles around the central supermassive black holes and consistently accounting for attenuation effects during propagation through the Earth. The dark matter--nucleon scattering cross section is evaluated by including elastic, resonant single pion production, and deep inelastic contributions, with particular emphasis on resonant single-pion production channels in order to smoothly cover the transition between the elastic and deep inelastic regimes. Using \IC neutrino data, we derive constraints on the parameter space of the model and show that this detection strategy can surpass the sensitivity of conventional direct-detection experiments for dark matter masses below $\sim 1$ GeV. We find that the signal is dominated by deep inelastic scattering and is therefore more sensitive to comparatively heavy mediators, while resonance processes provide a reduction of the event rate, reaching up to about $9\%$ near the experimental threshold. Our results demonstrate that \IC constitutes a powerful probe of sub-GeV dark matter scenarios through the observation of blazar-boosted dark matter.
\end{abstract}
\preprint{}
\maketitle


\section{Introduction}

Dark matter (DM) is a key ingredient in Nature, as indicated by the various gravitational signatures pointing towards its existence, which the Standard Model (SM) cannot explain. In the last decades, a considerable effort has been taken to detect this hidden component of the Universe (see for example~\cite{Cirelli:2024ssz}), focusing mainly on three branches: (1) DM production at colliders, (2) DM indirect detection due to annihilation or decay into SM particles, and (3) DM direct detection through the scattering of incident DM particles off targets in terrestrial detectors. However, all of these experiments have unfortunately yielded null results.

From the point of view of direct detection experiments, searches are limited to masses above the GeV scale, since a large recoil energy is required for DM-nucleon scattering. This has motivated the scientific community to explore new scenarios featuring sub-GeV DM that could be probed via direct detection. One of these possibilities comes from boosted DM, which occurs when energetic secondary particles transfer their energy to otherwise non-relativistic DM particles. Given this interaction, the boosted DM can now scatter with nucleons at detectors, depositing a measurable amount of energy. Various sources of acceleration have been studied in the literature, for example, cosmic ray boosted DM (CRBDM)~\cite{Bringmann:2018cvk, Dent:2019krz, Lei:2020mii, Super-Kamiokande:2022ncz, Maity:2022exk, Wang:2023wrx, Bell:2023sdq, PandaX:2024pme, Cappiello:2024acu, Ghosh:2024dqw, Diurba:2025lky, LZ:2025iaw,Yu:2025bwu}, boosted DM from semi-annihilations or decays (DABDM)~\cite{Kopp:2015bfa, Necib:2016aez, Fornal:2020npv, Toma:2021vlw, Aoki:2023tlb, BetancourtKamenetskaia:2025noa, Clark:2024baf}, neutrino boosted DM ($\nu$BDM)~\cite{Yin:2018yjn, Jho:2021rmn, Das:2021lcr, Lin:2022dbl, Lin:2023nsm, Lin:2024vzy, DeRomeri:2023ytt, Das:2024ghw, Ghosh:2024dqw, Sun:2025gyj} and blazar boosted DM (BBDM)~\cite{Wang:2021jic, Granelli_2022ysi, Bhowmick:2022zkj, DeMarchi:2024riu, CDEX:2024qzq, Gustafson:2025dff, Jeesun:2025gzt, Wang:2025ztb, Dev:2025czz, DeMarchi:2025uoo, Barillier:2025xct}.

The aim of this work is to explore the detection of BBDM in the GeV and sub-GeV sector. This kind of boosted DM is sourced by some of the most energetic environments observed in astrophysics: particle jets accelerated in the vicinity of AGNs and oriented roughly toward Earth. Interestingly, the DM density around a blazar is enhanced by the strong gravitational field of the supermassive black hole at its center, forming a density spike on top of the underlying halo distribution~\cite{Gondolo:1999ef, Merritt:2003qc, Sadeghian:2013laa}. The combination of a large DM target density with the extremely energetic particle spectrum makes blazars a particularly powerful setting for boosted DM searches, leading to non-negligible fluxes with energies larger than $\sim 1$~TeV (which is non-trivial for CRBDM). These large energies are precisely what allows the boosted DM to be detected, in spite of its relatively small mass.

This scenario has the advantage of being possible through simple requirements, as its boosting and detection relies only on interactions with nucleons. In contrast, $\nu$BDM depends on additional interactions with neutrinos (which in turn have constraints~\cite{Cline_2022,Ferrer:2022kei,Cline:2023tkp,Zapata:2025huq}), while DABDM hinges on the existence of a complex dark sector. An additional advantage of this scenario is that observational data is available to model the resulting BBDM flux. In particular, Ref.~\cite{Rodrigues:2023vbv} provides proton and electron fluxes for 324 AGNs, inferred from multiwavelength observations. While this sample is not exhaustive relative to the full CGRaBS catalog~\cite{Healey:2007gb} from which it is drawn\footnote{For more information, visit: \url{https://heasarc.gsfc.nasa.gov/W3Browse/radio-catalog/cgrabs.html}}, it nevertheless offers a representative estimate of the expected total BBDM contribution.

Currently, limits in boosted DM have already been established. The first paper on the subject~\cite{Bringmann:2018cvk} used data from MiniBooNE and XENON1T to constrain the spin-independent non-relativistic DM-nucleon cross section using CRBDM. This cross section was bounded down to about $10^{-31}$~cm$^2$, for DM masses as low as 100~keV. Recently, the LZ experiment has placed strong bounds on this scenario~\cite{LZ:2025iaw}, restricting cross sections as low as $10^{-33}$~cm$^2$, for keV DM masses. Furthermore, in~\cite{Cappiello:2024acu} the prospects of detecting CRBDM at \IC was explored, expecting to exclude cross sections in the ballpark of $10^{-41}$~cm$^2$ if DM masses are around 1~keV. From the point of view of BBDM, in~\cite{Wang:2021jic} the authors use XENON1T and MiniBooNE to constrain the non-relativistic cross section using data from TXS~0506+056 and BL~Lacerta. For masses as low as 10~keV, they rule out cross sections of order $10^{-36}$~cm$^2$. Similar bounds have been obtained by the CDEX-10 experiment~\cite{CDEX:2024qzq}. Morover, in~\cite{Wang:2025ztb} the importance of including both elastic and deep-inelastic scattering (DIS) interactions has been argued, as well as the need to go to model-dependent scenarios to provide more reliable results, setting thus stringent bounds for BBDM at \ICns.

In this work we intend to assess the sensitivity of \IC to BBDM, within a fermionic DM model provided with a vector mediator. One of the novelties of the present analysis is the explicit evaluation of resonant single pion production, which provides a more complete description of the transition between the elastic and DIS regimes.  Furthermore, we carry out the calculation in the context of conservative models for the DM spike profile around the blazar, which has not always been the case in the literature. Finally, instead of concentrating on specific blazars, we perform a scan over the 324 objects in~\cite{Rodrigues:2023vbv}, providing a sort of diffuse DM flux reaching the Earth.\footnote{Note that after attenuation effects only a handful of blazars will be responsible for the events measured at the detector.} We concentrate our efforts of detecting this flux at \ICns, as its kilometer-scale instrumented volume, excellent energy resolution for cascade-like events, and low background at high energies make it particularly well suited for such searches.

The paper is organized as follows. In \Cref{sec:cross_sections} we introduce the interaction model and the corresponding cross sections employed to derive the DM fluxes. In \Cref{sec:1_th_frame}, we develop the framework used to compute DM fluxes from blazars. \Cref{subsec:1a_proton_flux} describes the calculation of proton fluxes by introducing a blob frame in which the particle distributions are isotropic. In \Cref{subsec:1b_DM_rho}, we present the model for the DM density around AGNs, including a spike component superimposed on a Navarro–Frenk–White profile~\cite{Navarro:1995iw, Navarro:1996gj}. In \Cref{subsec:1d_DM_flux}, we outline the formalism used to obtain the DM fluxes from the previously computed quantities and derive the final spectra for the interactions considered. We conclude \Cref{sec:1_th_frame} with \Cref{subsec:1e_attenuation}, where we discuss the attenuation of the flux as it propagates through the Earth before reaching \ICns. In \Cref{sec:4_IC}, we present the simulation results, using \IC to constraint the non-relativistic cross section of the model, for the various benchmarks and scenarios explored. Finally, our conclusions are given in \Cref{sec:4_conclusions}.


\section{Model and Cross-Sections}
\label{sec:cross_sections}

For definiteness, we will consider fermionic DM $\chi$ which interacts with the SM exclusively through a dark vector mediator, which we will refer to as $Z_\mu^\prime$. The latter has vector couplings with both DM and SM particles. The general form that the Lagrangian takes is the following,
\begin{equation}
\begin{split}
\mathcal{L} = g_D\, \overline{\chi}\, \gamma^\mu \chi \,Z^\prime_\mu + g_{NZ^\prime}\sum_q \overline{q}\, \gamma^\mu q \,Z^\prime_\mu\,,
\end{split}
\end{equation}
where $q$ runs only over quarks, and $g_D$ and 
$g_{NZ^\prime}$ are the $Z'$ couplings to DM and to each quark of the SM, respectively. For simplicity, we are not including couplings to leptons, since we will consider the main contribution to the boosting and detection mechanisms to come from interactions with nucleons\footnote{Notice that couplings of electrons to dark vectors can be strongly constrained by the NA64 experiment~\cite{Banerjee:2019pds,NA64:2023wbi,NA64:2025ddk}. Instead, there can be constraints coming from meson decays (see for instance \cite{NA62:2025upx}). However, in our case the vector coefficients to quarks are all the same, so we expect interactions with mesons to be negligible, since the vector couplings to quarks and antiquarks are of opposite sign.}.

In this work, we have selected different cases for our study. First, we test five different DM masses in the GeV / sub-GeV mass range, usually under the reach of traditional direct detection experiments: $m_\chi = 1,\, 10,\, 100$ MeV, and $1,\, 10$ GeV. Furthermore, for the mediator we consider three different masses: $m_{Z'} = 1$ GeV, $m_{Z'} = 10$ MeV, and $m_{Z'} = 3 \, m_\chi$.

In the following, we describe how we model the three contributions to the DM - nucleon cross section: elastic interactions, single pion resonant production (RES), and deep inelastic scattering (DIS). In practice, the most important process for the energies under consideration is DIS but, as commented in the Introduction, the former two can provide non-negligible corrections.

\subsection{Elastic cross section}

When computing the DM-nucleon elastic cross section for the process $\chi (k_1) + N(p_1) \to \chi (k_2) + N(p_2)$, one finds the amplitude:
\begin{equation}
\begin{split}
\mathcal{M} = \frac{g_D \,g_{NZ^\prime}}{q^2 - m_{Z^\prime}^2} \bar{u}_\chi(k_2) \gamma^\mu u_\chi (k_1) \langle N(p_2) | V^{Z^\prime Q}_\mu (0) | N(p_1) \rangle \,,
\end{split}
\end{equation}
where $V^{Z^\prime Q}_\mu$ is the full hadronic current for the interaction with the $Z'$, which is purely vectorial in our case. To calculate the matrix elements, we relate them to form factors:
\begin{equation}
\begin{split}
\langle N(p_2) | V^X_\mu | N(p_1) \rangle &= \bar{u}_N(p_2) \left[ \gamma_\mu F_1^{XN}(Q^2) + \frac{i\sigma_{\mu\nu}q^\nu}{2m_N} F_2^{XN}(Q^2) \right] u_N(p_1)\,,
\end{split}
\end{equation}
where $V^X_\mu$ are generic vector operators, and $F_1^{XN}(Q^2)$, and $F_2^{XN}(Q^2)$ are the corresponding Dirac, and Pauli form factors, respectively. Moreover, $q^\mu\equiv (p_2-p_1)^\mu$, $Q^2=-q^2$, and $m_N$ is the nucleon mass. Here we are considering no contributions coming from second-class currents, so that our BSM interactions are G-invariant~\cite{Weinberg:1958ut}.

To get the form factors, we can use the parametrization in~\cite{Bradford:2006yz,Bodek:2007ym},
which is a modification of the Kelly form factors~\cite{Kelly:2004hm}. We refer the reader to the cited works for more details about their method. After some algebraic computations, we can decompose the vectorial currents of the dark sector model in terms of the SM ones, by neglecting contributions of the $c$, $b$ and $t$ quarks:
\begin{equation}
\begin{split}
V^{Z^\prime Q}_\mu &= -6\,V^3_\mu + 6\, J^\mathrm{EM}_\mu + 3\,V^s_\mu \,,
\end{split}
\end{equation}
where $V^3_\mu \equiv \frac{1}{2} \left[\bar{u} \gamma_\mu u - \bar{d} \gamma_\mu d \right]$ is the vectorial current of the third component of isospin $I_3$, $J^\mathrm{EM}_\mu \equiv \frac{2}{3}\sum_\alpha\left[\bar{q}^U_\alpha \gamma_\mu q^U_\alpha \right] - \frac{1}{3} \sum_\alpha \left[\bar{q}^D_\alpha \gamma_\mu q^D_\alpha \right]$ is the electromagnetic current, with $q^U_\alpha,\,q_\alpha^D$ the $u,d\,$-like quarks respectively, and $V^s_\mu \equiv \bar{s} \gamma_\mu s$ is the strange vectorial current.

By using these relations between the $Z'$ induced currents and the SM ones, we can obtain our form factors in terms of the SM ones:
\begin{equation}
\begin{split}
F_{1,2}^{Z^\prime N} &= \mp 3\left(F_{1,2}^p - F_{1,2}^n \right) + 6\, F_{1,2}^\mathrm{EM, N} + 3\, F_{1,2}^{sN} \,,
\end{split}
\end{equation}
where all SM form factors are modeled in~\cite{Bradford:2006yz,Bodek:2007ym}. With these, the amplitude can be computed and thus the elastic cross section, $\sigma_{\rm el}$. We do not write the final result explicitly due to its length.

\subsection{Single pion resonant production (RES)}

Another relevant contribution to the total cross section between dark matter fermions and nucleons comes from single pion production through $\Delta-$ and $N-$resonances, $\chi(k_1) + N(p_1) \to \chi(k_2) + N^*(p_2) \to \chi(k_2) + N(p_3) + \pi(p_4)$. These inelastic processes are not fully captured by the modeling of deep inelastic cross sections, so they need to be included separately. For astrophysical scenarios and when we are just interested in counting the total number of events, it is enough to work from the Feynman, Kislinger and Ravndal model (FKR)~\cite{Feynman:1971wr} to compute the transitions between different helicity states of nucleons to the resonances, such as was performed by Ravndal~\cite{Ravndal:1971cuf} and more generally by Rein and Sehgal~\cite{Rein:1980wg} for neutrino induced single pion production. This approach was later expanded to include matter effects~\cite{Kuzmin:2003ji, Berger:2007rq, Graczyk:2007bc}. We will follow the approach by \cite{HoefkenZink:2025tns}, suitable for DM interactions and built upon the Rein and Sehgal neutrino model.

If we consider protons and neutrons, there are four possible channels:
\begin{enumerate}
    \item $\chi + p \to \chi + p + \pi^0$,
    \item $\chi + p \to \chi + n + \pi^+$,
    \item $\chi + n \to \chi + n + \pi^0$,
    \item $\chi + n \to \chi + p + \pi^-$.
\end{enumerate}
The amplitude $\mathcal{M}_r$ for a process where a resonance $r$ is created reads~\cite{HoefkenZink:2025tns}:
\begin{multline}
\label{eq:M_vec}
\mathcal{M}_r(\chi(k_1, \lambda_1) N(p_1) \to \chi(k_2, \lambda_2) N^{*} (p_2))=\\
2W \frac{g_D\, g_{NZ'}}{q^2 - m^2_{Z^\prime}} \Big( \bra{N^{*}_r} C_L^{\lambda_1 \lambda_2} F_-^V + C_R^{\lambda_1 \lambda_2} F_+^V + C^{\lambda_1 \lambda_2}_{0} F_0^{V, \lambda_1 \lambda_2} \ket{N} \Big)\,,
\end{multline}
where the coefficients $C^{\lambda_1 \lambda_2}_i$ are the left ($L$), right ($R$) and longitudinal (0) polarization components of the matrix element $\left[\bar{u}_{\chi ,\lambda _2} \!\gamma _{\mu } u_{\chi ,\lambda _1} \right] \! \left(\!g^{\mu \nu }- \! \frac{q^{\mu} q^{\nu} }{M_{Z'}^2}\right)$, with $\lambda_k$ being the helicities of the DM fermions. The $F^{V}_i$ are the vector currents operators defined in the FKR model (not to be confused with form factors), and $W \equiv \sqrt{p_2^2}$ is the invariant mass of the outgoing nucleon and pion.  With this amplitude, we can compute the differential cross sections,
\begin{equation}
\label{eq:diff_cs_v-a}
\begin{split}
\frac{d^2 \sigma_{\rm RES}}{dq^2 dW} = \frac{g_D^2\, g_{NZ'}^2}{4 \pi^2\left(q^2 - m^2_{Z^\prime}\right)^2} \frac{W}{m_N} \sum_{\lambda_1 \lambda_2} \left( \left| C_L^{\lambda_1 \lambda_2} \right|^2 \sigma_L + \left| C_R^{\lambda_1 \lambda_2} \right|^2 \sigma_R + \left| C_0^{\lambda_1 \lambda_2} \right|^2 \sigma_0^{\lambda_1 \lambda_2} \right)\,,
\end{split}
\end{equation}
where the helicity-cross sections $\sigma_i^{(\lambda_1 \lambda_2)}$ are expressed as follows:
\begin{equation}
\begin{split}
\sigma_R &= \frac{\pi}{16} \frac{W}{m_N |\vec{p}_{\mathrm{IB},1}|^2} \sum_{\mathcal{C}_R}\sum_{j_z} \left|\sum_{r \in \mathcal{C}_R}  \mathcal{F}_R \times \bra{N, j_z + 1} F_+^V \ket{N^*_r, j_z}  \right|^2,\\[1mm]
\sigma_L &= \frac{\pi}{16} \frac{W}{m_N |\vec{p}_{\mathrm{IB},1}|^2} \sum_{\mathcal{C}_R}\sum_{j_z} \left|\sum_{r \in \mathcal{C}_R} \mathcal{F}_R \times \bra{N, j_z - 1} F_-^V \ket{N^*_r, j_z} \right|^2,\\[1mm]
\sigma_0^{\lambda_1 \lambda_2} &= \frac{\pi}{16} \frac{W}{m_N |\vec{p}_{\mathrm{IB},1}|^2} \sum_{\mathcal{C}_R}\sum_{j_z} \left|\sum_{r \in \mathcal{C}_R} \mathcal{F}_R \times \bra{N, j_z} F_{0^\pm }^{V, \lambda_1 \lambda_2} \ket{N^*_r, j_z} \right|^2,
\end{split}
\end{equation}
with $\vec{p}_{\mathrm{IB},1}$ the 3-momentum of the incoming $\chi$ in the isobaric frame~\cite{Adler:1968tw}, where the resonance is at rest, and $\mathcal{C}_R \equiv \{ S_{r 1}, P_{r 1}, P_{r 3}, D_{r 3}, D_{r 5}, F_{r 5}, F_{r 7} \}$
represents the groups of resonances that need to be summed coherently.\footnote{Here we group each resonance $r$ in terms of orbital angular momentum $L$ and total angular momentum $J$: $L_{r\,2J}$. There are 18 resonances in total (and up to 6 possible different spin transitions for each DM-helicity combination), with resonances with the same $L$ and $J$ capable of interference. These are summed over different isospin states.} In addition, the factor
\begin{equation}
\begin{split}
\mathcal{F}_R \equiv \mathrm{sgn} (N^*_r) \sqrt{\frac{1}{2\pi N_r} \frac{\Gamma_r}{\left( W - M_r \right)^2 + \Gamma_r^2 / 4} \left(x_E^r \right)^2}\,,
\end{split}
\end{equation}
accounts for the nearly on-shell production of the baryon resonances, with mass $M_r$.
Here, $N_r$ is a normalization factor, $\Gamma_r$ is the resonance decay width, $x_E^r$ is the resonance elasticity and $\mathrm{sgn} (N^*_r)$ is the sign of the amplitude of the decay, important for interferences between different mediating resonances.

The structures above are related to the helicity amplitudes:
\begin{equation}
\begin{split}
f_{\pm |2j_z|}^{V} &\equiv \bra{N, j_z \pm 1} F_\pm^{V} \ket{N^*_r, j_z},\\[1mm]
f_{0^\pm }^{V, \lambda_1 \lambda_2} &\equiv \bra{N, \pm 1/2} F_{0^\pm }^{V, \lambda_1 \lambda_2} \ket{N^*_r, \pm 1/2}.
\end{split}
\end{equation}
which are calculated using the FKR model. For more details about the computation of $f_{\pm |2j_z|}^{V}$ and $f_{0^\pm }^{V(A), \lambda_1 \lambda_2}$, we refer the reader to \cite{HoefkenZink:2025tns}. The total differential cross section is found by summing the processes mediated by each of the $18$ baryon resonances, following the interference rules found in \cite{Rein:1980wg}.

\subsection{Deep inelastic scattering (DIS)}

At the highest energies, the most important contribution comes from deep inelastic scattering (DIS), which is the process for which the interaction with the constituents of the nucleons can be computed with the parton model, $\chi (k_1) + N(p_1) \to \chi (k_2) + X(p_2)$. As can be seen in \cite{HoefkenZink:2024hor}, since we have purely vector couplings, the differential cross section of the process in the lab frame of the nucleons is 
 \begin{equation}
 \label{eq:DIS}
 \begin{split}
 \frac{d^2\sigma_{\rm DIS}}{dx\,dy} = & \frac{g_D^2\, g_{N Z^\prime}^2}{4 \pi m_{Z^\prime}^4} \frac{E_\chi^2\, m_N x}{\, (1 + Q^2 / m_{Z^\prime}^2 )^2} \frac{\sqrt{E_\chi^2 (1 - y)^2 - m_\chi^2}}{(1-y)(E_\chi^2 - m_\chi^2)} \left( \left(y^2 - 2y + 2 \right) - \frac{ m_\chi^2}{E_\chi m_N x} y \right) \sum_q f_q(x, Q^2)\,,
 \end{split}
 \end{equation}
 where $f_q(x, Q^2)$ is the parton distribution function for the corresponding quark, $E_\chi$ and $m_\chi$ are the energy and mass of the incoming DM fermion, $x \equiv \frac{Q^2}{2 p_1\cdot q}$ is the Bjorken scaling variable, and $y \equiv \frac{p_1\cdot q}{p_1\cdot k_1}$ is the inelasticity.

\subsection{Numerical Results}

For illustration, in the following we shall present total DM-proton cross sections $\sigma_{\chi p}$ as a function of $E_\chi$, representing the incoming DM energy when the nucleon is at rest. The cross sections for neutrons are very similar. These results are particularly relevant in order to understand the attenuation and detection of the boosted DM flux.

\begin{figure}
    \centering
    \includegraphics[width=0.49\linewidth]{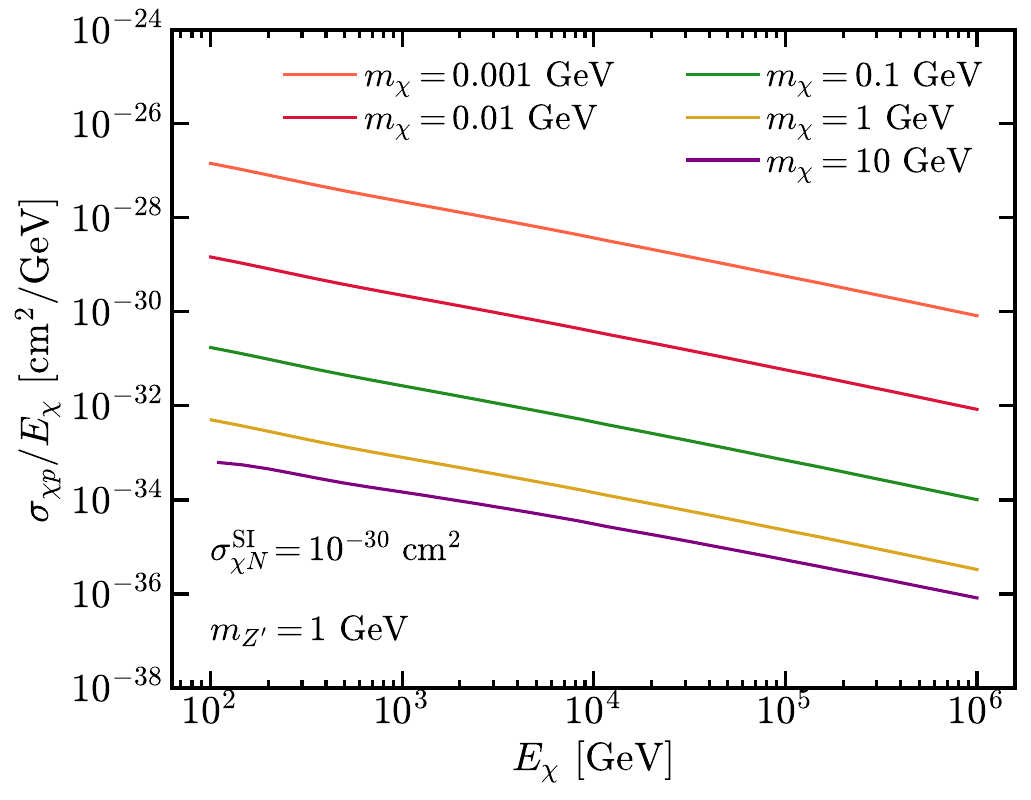}
    \includegraphics[width=0.49\linewidth]{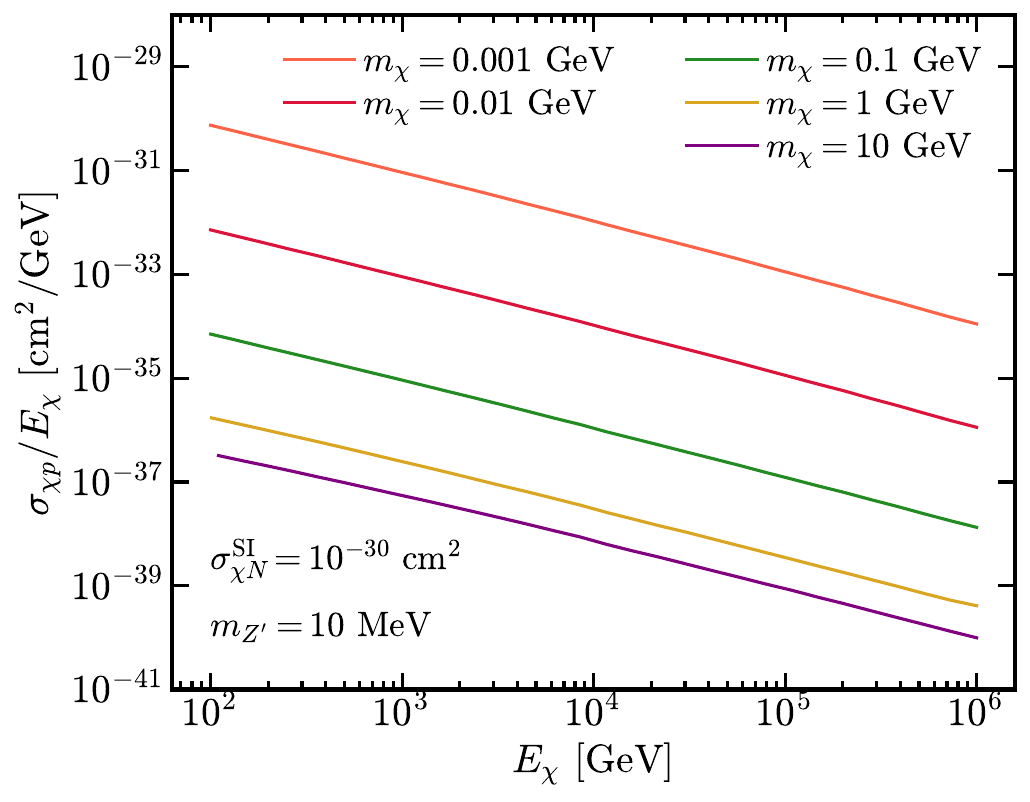}
    \includegraphics[width=0.49\linewidth]{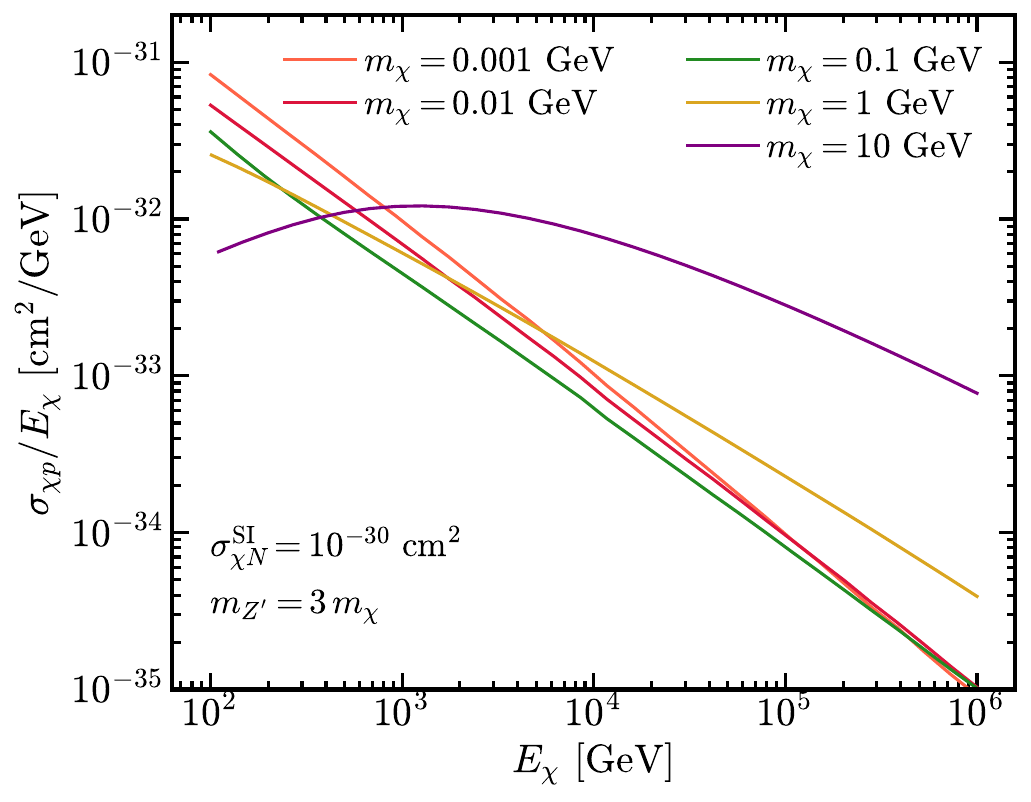}
    \caption{Total cross sections for each of the cases selected, with the respective non relativistic total cross section fixed to $\sigma^\mathrm{SI}_{\chi N} = 10^{-30} \, \mathrm{cm}^2$. We are only considering energies relevant for high energy events at \IC.}
    \label{fig:cs_NR}
\end{figure}

For energies relevant for \ICns, $E_\chi \gtrsim 10^2$~GeV, we find that it is usually the tail of the $\sigma_{\chi p}/E_\chi$ distribution which affects our observables, explained largely by DIS, with corrections from RES and elastic scattering of $3$ and $4$\% at most, respectively. 
The total cross section as a function of $E_\chi$ is shown in Figure~\ref{fig:cs_NR} for different scenarios. For each curve, the couplings are fixed such that the non-relativistic cross section $\sigma^{\rm SI}_{\chi N}$ is $10^{-30}\, \mathrm{cm}^2$ (see Appendix~\ref{sssec:non_rel_cs}).

In each of the upper panels of the Figure, the mediator mass is set at a specific value. We find that, for a given mediator mass, larger DM masses correspond to smaller cross sections, with the variation spanning several orders of magnitude. This separation of the curves along the vertical axis can be largely
understood from the reduced-mass dependence induced by fixing
$\sigma^{\rm SI}_{\chi N}$. Indeed, from \Cref{eq:NR_CrossSection}, one finds that
$g_D^2 \, g_{NZ'}^2/m_{Z'}^4 \propto \mu_{\chi N}^{-2}$, so that the overall
normalization of the high-energy cross section is largely controlled by
$\mu_{\chi N}^{-2}$. For a fixed mediator mass, increasing
$m_\chi$ increases $\mu_{\chi N}$ and therefore requires a smaller effective
coupling combination $g_D g_{NZ'}$. In particular, for $m_\chi \ll m_N$,
$\mu_{\chi N}\simeq m_\chi$, and the dependence approaches $m_\chi^{-2}$,
explaining the large vertical separation between the curves for the lighter
DM masses.

Moreover, comparing the two upper panels, we can also conclude that, for a given DM mass, increasing $m_{Z'}$ increases the cross section. In this case, a modification of the mediator mass leads to larger couplings in \Cref{eq:NR_CrossSection}; however, this effect is canceled in \Cref{eq:DIS} for DIS. The reason for this increase actually lies in the $(1+Q^2/m_{Z'}^2)^{-2}$ factor, which increases as $\sim m_{Z'}^4$ for large $Q^2$.

The lower panel of the Figure considers the case when $m_{Z'}=3m_\chi$, such that for all curves the mediator mass is always heavier than the DM mass. Here we should not expect a situation similar to the other panels, as each curve has a different mediator mass. In general, we see that for low (large) $E_\chi$ the cross section decreases (increases) with $m_\chi$. This reversal of ordering happens for two reasons. First, one finds that the peak of the distribution is smaller for larger $m_\chi$. Second, for larger $m_\chi$ (and thus $m_{Z'}$) the peak happens at larger values of $E_\chi$, leading to a comparatively smaller slope. In general, the peak of the distribution is placed at $E_\chi<10^2$~GeV, shifting towards larger energy as the mediator mass increases. However, for $m_{\chi}=10$~GeV, which has $m_{Z'}=30$~GeV, one can see the smaller peak at $E_\chi\approx10^3$~GeV, with a much gentler slope than the other cases at larger energies.


\section{Dark Matter Flux from Blazars}
\label{sec:1_th_frame}

\subsection{Proton flux}
\label{subsec:1a_proton_flux}

In order to describe the flux of protons emitted by a blazar, one considers an outgoing jet, inclined with respect to the observer by an angle $\theta_\mathrm{LOS}$. Here, the analysis is focused on a homogeneous region following the jet, referred to as the blob~\cite{Dermer:2009zz}, which propagates at speed $\beta_\mathrm{B}$ from the origin of the blazar emission.

Jet particles at the rest frame of this blob are distributed isotropically. For (lepto-)hadronic models, the proton energy spectrum in this frame follows a power law distribution, with power $\alpha_p$~\cite{kardashev1962nonstationarity, Cerruti:2020lfj}
\begin{eqnarray}
\frac{d \Gamma'_p}{d E'_p\, d \Omega'} = \frac{1}{4 \pi} c_p \left( \frac{E'_p}{m_p} \right)^{-\alpha_p}\,,
\end{eqnarray}
where 
$m_p \simeq 0.938$ GeV is the proton mass,
and $c_p$ is a normalization constant determined from the proton luminosity $L_p$, such that: 
\begin{equation}
c_p = \frac{L_p}{m_p \,\Gamma_\mathrm{B}^2 \int_{\gamma^\prime_{\min}}^{\gamma^\prime_{\max}} dx \ x^{1 - \alpha_p}}\,,
\end{equation}
with $\Gamma_B = \sqrt{1-\beta_B^2}$ the Lorentz boost factor of the blob, and $\gamma^\prime_{\min / \max}$ are the minimum / maximum boost factors of the protons, $\gamma^\prime_{\min} \leq E_p^\prime / m_p \leq \gamma^\prime_{\max}$.

As shown in \cite{Wang:2021jic}, in the reference system of the observer the proton spectrum is
\begin{eqnarray}
\label{eq:ProtonSpectrum}
\frac{d \Gamma_p}{d T_p \,d \Omega} = \frac{1}{4 \pi} c_p \left( 1+\frac{T_p}{m_p} \right)^{-\alpha_p} \frac{\beta_p(1-\beta_p\, \beta_B\, \mu)^{-\alpha_p} \Gamma_B^{-\alpha_p}}{\sqrt{(1-\beta_p\, \beta_B\, \mu)^2-(1-\beta_p^2)(1-\beta_B^2)}}\,,
\end{eqnarray}
where $T_p = E^\prime_p - m_p$ is the proton kinetic energy, $\mu$ is the cosine of the angle between each proton and the jet (see \Cref{app:kinematic_rels}), and $\beta_p = \sqrt{1-m_p^2/(T_p+m_p)^2}$ is the proton speed. The parameters $\alpha_p,\,\beta_B,\,\gamma'_{\rm min},\,\gamma'_{\rm max},\,L_p$ for each blazar considered in this work are taken from~\cite{Rodrigues:2023vbv}.

\subsection{Dark matter density profile}
\label{subsec:1b_DM_rho}

To model the dark matter distribution around a blazar, we follow the same treatment as~\cite{Wang:2021jic, Granelli_2022ysi, DeMarchi:2024riu, Wang:2025ztb, DeMarchi:2025uoo}. This distribution should form a spike close to the supermassive BH, exhibiting a steeper slope than that of the halo. If the halo density follows a power law, $\rho_\mathrm{DM}^h = \mathcal{N}\, r^{- \gamma}$, where $\mathcal N$ is a normalization constant and $r$ is the distance to the center of the BH, then the full density distribution of DM around the supermassive BH is~\cite{Gondolo:1999ef, DeMarchi:2025uoo}
\begin{equation}
\rho_\mathrm{DM} (r) = \mathcal{N}\, R_\mathrm{sp}^{- \gamma} \begin{cases}
    0, & \text{if } r \leq 2 R_s\,, \\
    \left( 1 - \frac{2 R_s}{r}  \right)^{3/2} \left( \frac{R_\mathrm{sp}}{r} \right)^{\frac{9 - 2 \gamma}{4 - \gamma}}, & \text{if } 2 R_s \leq r \leq R_\mathrm{sp}\,, \\
    \left( \frac{R_\mathrm{sp}}{r} \right)^\gamma, & \text{if } R_\mathrm{sp} \leq r\,,
\end{cases} \,,
\end{equation}
where $R_\mathrm{sp}$ is the radial distance up to which the spike extends and $R_s$ is the Schwarzschild radius. The radius of the spike can be approximated as~\cite{Merritt:2003qc},
\begin{equation}
R_\mathrm{sp} \simeq 0.1 \left( \frac{M_\mathrm{BH}}{\mathcal{N}} \right)^{\frac{1}{3 - \gamma}}\,,
\end{equation}
where $M_\mathrm{BH}$ is the mass of the BH. In the region of the spike we included a factor $\left( 1 - \frac{2 R_s}{r}  \right)^{3/2}$ that accounts for the capture of DM onto the BH~\cite{Sadeghian:2013laa}. In the following, we will consider a Navarro-Frenk-White distribution for the halo~\cite{Navarro:1995iw, Navarro:1996gj}, such that $\gamma = 1$ and $\mathcal{N} \simeq 10^{-14} M_\mathrm{BH} / R_s^2$, with $M_\mathrm{BH}$ for each blazar taken from~\cite{Rodrigues:2023vbv}.

In order to compute the boosted DM flux, we need to calculate the radial integral of the density up to the spike, $\Sigma_\mathrm{DM}^s\equiv \Sigma_\mathrm{DM} (R_\mathrm{sp})$, where the LOS (line-of-sight) integral is, 
\begin{equation}
\label{eq:ColumnDensity}
\Sigma_\mathrm{DM} (r) = \int_{r_{\min}}^r \rho_\mathrm{DM} (r') dr'\,.
\end{equation}
with $r_{\min}<r<R_{\rm sp}$ the region where DM is accelerated. For our final results, we define two benchmark points related to $r_\mathrm{min}$. Following~\cite{DeMarchi:2025uoo}, we define our first benchmark, BP1, as $r_{\min} = 100 \, R_s$. Another suitable choice for a benchmark would be the dissipation radius~\cite{Rodrigues:2023vbv}, however, the catalog does not contain this value for all of the blazars we are considering. As an alternative, since this radius is of the same order of $R_{\rm BLR}$, the radius for the broad line region, we define our second benchmark, BP2, by $r_{\min} = R_\mathrm{BLR}$.

The DM distribution may also be altered by annihilation processes~\cite{Gondolo_1999, Shapiro:2016ypb}, which will not be considered in our work.

\subsection{Dark matter flux from blazars}
\label{subsec:1d_DM_flux}

For any blazar, we can calculate the proton flux within its jet from Eq.~(\ref{eq:ProtonSpectrum}) and the DM column density around it from Eq.~(\ref{eq:ColumnDensity}). Then the boosted DM flux arriving to Earth from a specific blazar~\cite{Wang:2021jic, DeMarchi:2025uoo} has three main components, depending on the type of interaction:
\begin{equation}
\label{eq:BoostedFlux1}
\begin{split}
\left(\frac{d \Phi_{\chi}}{d T_{\chi}}\right)_\mathrm{EL} &= \frac{\Sigma_\mathrm{DM}^{s} }{2 \pi\, m_{\chi}\, d_{L}^2} \int_0^{2 \pi} d \phi_s \int_{T_{p}^{\min} \left(T_\chi, \phi_s\right)}^{T_{p}^{\max} \left(T_\chi, \phi_s\right)} d T_p\, \frac{d\sigma_\mathrm{el}}{dQ^2}\frac{dQ^2}{dT_\chi} \times \frac{d \Gamma_p}{d T_p\, d \Omega} \left(T_p, \mu^\mathrm{EL} \right)\,,\\
\left(\frac{d \Phi_{\chi}}{d T_{\chi}}\right)_\mathrm{RES} &= \frac{\Sigma_\mathrm{DM}^{s} }{2 \pi\, m_{\chi}\, d_{L}^2} \int_0^{2 \pi} d \phi_s \int_{T_{p}^{\min} \left(T_\chi, \phi_s\right)}^{T_{p}^{\max} \left(T_\chi, \phi_s\right)} d T_p\, \int_{W_\mathrm{min}}^{W_\mathrm{max}} d W \frac{d\sigma^2_\mathrm{RES}}{dW dq^2} \left| \frac{\partial q^2}{\partial T_\chi} \right| \times \frac{d \Gamma_p}{d T_p\, d \Omega} \left(T_p, \mu^\mathrm{RES} \right) \,,\\
\left(\frac{d \Phi_{\chi}}{d T_{\chi}}\right)_\mathrm{DIS} &= \frac{\Sigma_\mathrm{DM}^{s} }{2 \pi\, m_{\chi}\, d_{L}^2} \int_0^{2 \pi} d \phi_s \int_{T_{p}^{\min} \left(T_\chi, \phi_s\right)}^{T_{p}^{\max} \left(T_\chi, \phi_s\right)} d T_p\, \int_{\nu_\mathrm{min}}^{\nu_\mathrm{max}} d \nu \frac{d\sigma^2_\mathrm{DIS}}{dx\, dy}  \left| \frac{\partial (x, y)}{\partial (T_\chi, \nu)} \right| \times \frac{d \Gamma_p}{d T_p\, d \Omega} \left(T_p, \mu^\mathrm{DIS} \right) \,,
\end{split}
\end{equation}
where $T_\chi$ is the outgoing kinetic energy of the boosted DM fermions at the blazar for an observer on Earth, and $d_{L}$ is the luminosity distance of each blazar. The angle $\phi_s$ is related to the solid angle in the proton jet as can be seen in \Cref{app:kinematic_rels} (see also Eq.~(40) in \cite{DeMarchi:2025uoo}). Furthermore, the limits of integration $T_p^{\min}$ and $T_p^{\max}$ for the proton kinetic energy $T_p$ are computed from the range of values of proton beam boost in the blob frame, $\gamma_\mathrm{min}^\prime$ and $\gamma_\mathrm{max}^\prime$, given by \cite{Rodrigues_2019} for each blazar, and are functions of $T_\chi$ and $\phi_s$.

The differential cross sections of the DM fermion with the protons have to be computed for the different contributions sketched in \Cref{sec:cross_sections}.
The Jacobians of the above processes are:
\begin{align}
\left| \frac{\partial q^2}{\partial T_\chi} \right| &= 2 m_\chi\,, &
\left| \frac{\partial (x, y)}{\partial (T_\chi, \nu)} \right| &= \frac{1}{\left(m_p+ T_p \right) \nu}\,.
\end{align}
where $\nu \equiv (p_1\cdot q)/m_N$ is the energy transfer of the process in the proton Lab frame, with $Q^2 = 2 m_\chi T_\chi$. 

Regarding the limits of integration for RES and DIS, we have $\nu_{\min} = m_\chi T_\chi / m_p$, as well as $\nu_{\max} = \min\left( m_\chi T_p / m_p\,, E_\chi^\mathrm{Lab} - m_\chi \right)$, and $E_\chi^\mathrm{Lab} = m_\chi \left( m_p + T_p \right) / m_p$. The limits of integration for the invariant mass $W$ are more elaborated, since the common limits are computed after integrating on $q^2$, and can be found in~\cite{HoefkenZink:2025tns}. Considering the process of boosting we are interested in, for a fixed $q^2$ there is an additional limit, such that the upper bound of integration in Eq.~(\ref{eq:BoostedFlux1}) should be the minimum between the one found in~\cite{HoefkenZink:2025tns} and: 
\begin{equation}
\begin{split}
W_{\max}^{q^2} = \sqrt{m_p^2 + q^2 + \frac{\left(m_p + T_p \right) \, q^2}{m_\chi} + \frac{\sqrt{ T_p \, q^2 \left( q^2 - 4 m_\chi^2 \right) \left(T_p + 2 m_p \right)} }{ m_\chi}}\,.
\end{split}
\end{equation}

Finally, the angles $\mu$ depend on the kinematical variables of the interaction under the regime considered, so that $\mu^\mathrm{EL} \equiv \mu^\mathrm{EL} \left( T_p, T_\chi, \phi_s \right)$, $\mu^\mathrm{RES} \equiv \mu^\mathrm{RES} \left( T_p, T_\chi, W, \phi_s \right)$, $\mu^\mathrm{DIS} \equiv \mu^\mathrm{DIS} \left( T_p, T_\chi, \nu, \phi_s \right)$. More details about these relations can be found in \Cref{app:kinematic_rels}.

It is important to keep in mind that the above cross sections are usually computed in the rest frame of the protons, while the boosting is naturally worked out in the rest frame of the DM.
The total fluxes would then be,
\begin{align}
\frac{d \Phi_{\chi}}{d T_{\chi}} = \left(\frac{d \Phi_{\chi}}{d T_{\chi}}\right)_\mathrm{EL} + \left(\frac{d \Phi_{\chi}}{d T_{\chi}}\right)_\mathrm{RES} + \left(\frac{d \Phi_{\chi}}{d T_{\chi}}\right)_\mathrm{DIS}\,.
\end{align}

Having defined all elements appearing in Eq.~(\ref{eq:BoostedFlux1}), one must now write the flux in terms of $T_{\chi}^{(0)}$, the kinetic energy measured on Earth after being redshifted~\cite{DeMarchi:2025uoo}:
\begin{equation}
\begin{split}
T_{\chi}^{(0)} = m_\chi \left( \sqrt{\frac{T_\chi \left( T_\chi + 2m_\chi \right)}{m_\chi^2 \left(1 + z \right)^2} + 1 } - 1 \right)\,,
\end{split}
\end{equation}
where $z$ is the redshift of the blazar. 
We have:
\begin{equation}
\begin{split}
\frac{d \Phi_{\chi}}{d T_{\chi}^{(0)}} = \frac{d \Phi_{\chi}}{d T_{\chi}} \times \frac{d T_{\chi}}{T_{\chi}^{(0)}}\,,
\end{split}
\end{equation}
where as in~\cite{DeMarchi:2025uoo}: 
\begin{equation}
\begin{split}
\frac{d T_{\chi}}{T_{\chi}^{(0)}} = \left(1 + z \right)^2 \frac{m_\chi + T_{\chi}^{(0)}}{\sqrt{m_\chi^2 + \left(1 + z \right)^2 \, T_{\chi}^{(0)} \left(2 m_\chi + T_{\chi}^{(0)} \right)}}\,.
\end{split}
\end{equation}

These calculations need to be performed for each of the 324 blazars under consideration by plugging in the fitted quantities for each of them in \cite{Rodrigues:2023vbv}, as indicated in Sections~\ref{subsec:1a_proton_flux} and~\ref{subsec:1b_DM_rho}.

It is worth noting that, for a given $\chi$ mass, the time of flight of each component of the total BBDM flux must be computed. If this time exceeds the age of the corresponding blazar, that component is removed from the spectrum, as the DM would not have had enough time to reach the Earth. We take the AGN age to be the minimum between 1 Gyr and the light-travel time to Earth. For the energy range relevant to our analysis, we find that this filter does not appreciably affect the spectrum, remaining essentially unchanged.

\subsection{Dark matter flux attenuation on Earth}
\label{subsec:1e_attenuation}

Our final objective is to obtain the number of DM events detected at the \IC detector, located at the South Pole at a depth of approximately $2$~km below the ice surface. However, before this can be achieved we need to calculate the attenuation of the DM flux as it traverses the Earth towards the detector, due to interactions of the former with nucleons within the planet.

The magnitude of this attenuation depends on the distance traveled inside the Earth, which is determined by the declination angle of the source AGN. Therefore, the attenuation of each component of the BBDM flux will differ for each AGN in our source list, with a minimum propagation through 2~km of ice, and the maximum reduction happening for AGNs located in the Northern sky, where the entire Earth has to be traversed.

The distance traveled by BBDM produced in a given AGN before reaching the detector can be computed as
\begin{eqnarray}
l_{\rm max}(\theta_{DE})= (R_{\oplus} - d) \sin{\theta_{DE}} 
+ \sqrt{(R_{\oplus} - d)^2 \sin^2{\theta_{DE}} - (R_{\oplus} - d)^2 +R_{\oplus}^2 }    
\end{eqnarray}
where $R_{\oplus} = 6\,371$ km is the radius of the Earth, $d \approx 2$ km is the average depth of \ICns, and $\theta_{DE}$ is the declination angle of the AGN. As an example, for the blazar TXS~0506+056 we have $\theta_{DE} = + 5.7$\textdegree, which leads to $l_{\rm max} \approx 1\,300$ km.

In addition to the distance $l_{\rm max}$, the attenuation of the BBDM flux is determined by the density of matter crossed by the DM particles along their path to the detector. This requires the Earth density profile $\rho(r)$, for which we adopt the one from the Preliminary Reference Earth Model (PREM)~\cite{Dziewonski:1981xy}. 
However, in order to use this density profile, we need to know the distance between each point of the DM trajectory and the center of the Earth, which can be written as
\begin{eqnarray}
r (l) = \sqrt{(R_{\oplus} - d)^2 + l^2 - 2 l (R_{\oplus} - d) \sin{\theta_{DE}}}
\end{eqnarray}

With both $l_{\rm max}$ and $\rho(r)$ we can determine the total amount of matter crossed by the DM flux, which is usually referred to as the column density $\Sigma_{\oplus}$. This is given by
\begin{eqnarray}
\Sigma_{\oplus} = \int_0^{l_{\rm max}} \rho(r(l)) \,dl~.
\end{eqnarray}
Moreover, the nucleon column density is defined as $\tau = \Sigma_{\oplus}/m_u$, with $m_u=931.494$~MeV being the atomic mass unit.
Therefore, for each of the 324 sources, after calculating $\phi_\chi$ with Eq.~(\ref{eq:BoostedFlux1}), the corresponding component of the attenuated BBDM flux at \IC is given by the cascade equation~\cite{Arguelles:2017atb}
\begin{equation}
\label{eq:cascade}
    \phi_{\chi}^{IC} (E_{\chi}) = \phi_{\chi}(E_{\chi}) ~\exp{ \left[-\tau(\theta_{DE}) ~  \sigma_{\chi N}(E_{\chi}) \right]} 
\end{equation}
where $E_\chi=T_\chi^{(0)}+m_\chi$, and $\sigma_{\chi N}(E_{\chi})$ is again calculated with 
the cross sections presented above (see Figure~\ref{fig:cs_NR})\footnote{Note that in \Cref{fig:cs_NR} we show the DM-proton cross section, while that appearing in \Cref{eq:cascade} corresponds to that for nucleons. The differences between cross sections for protons and neutrons are very small, so we take the average.}, and we have disregarded a subdominant term which redistributes high-energy DM particles to lower energies. Finally, the total flux at \IC is obtained by summing the contributions from all AGNs in the sample. 

\begin{figure}
    \centering
    \includegraphics[width=0.9\linewidth]{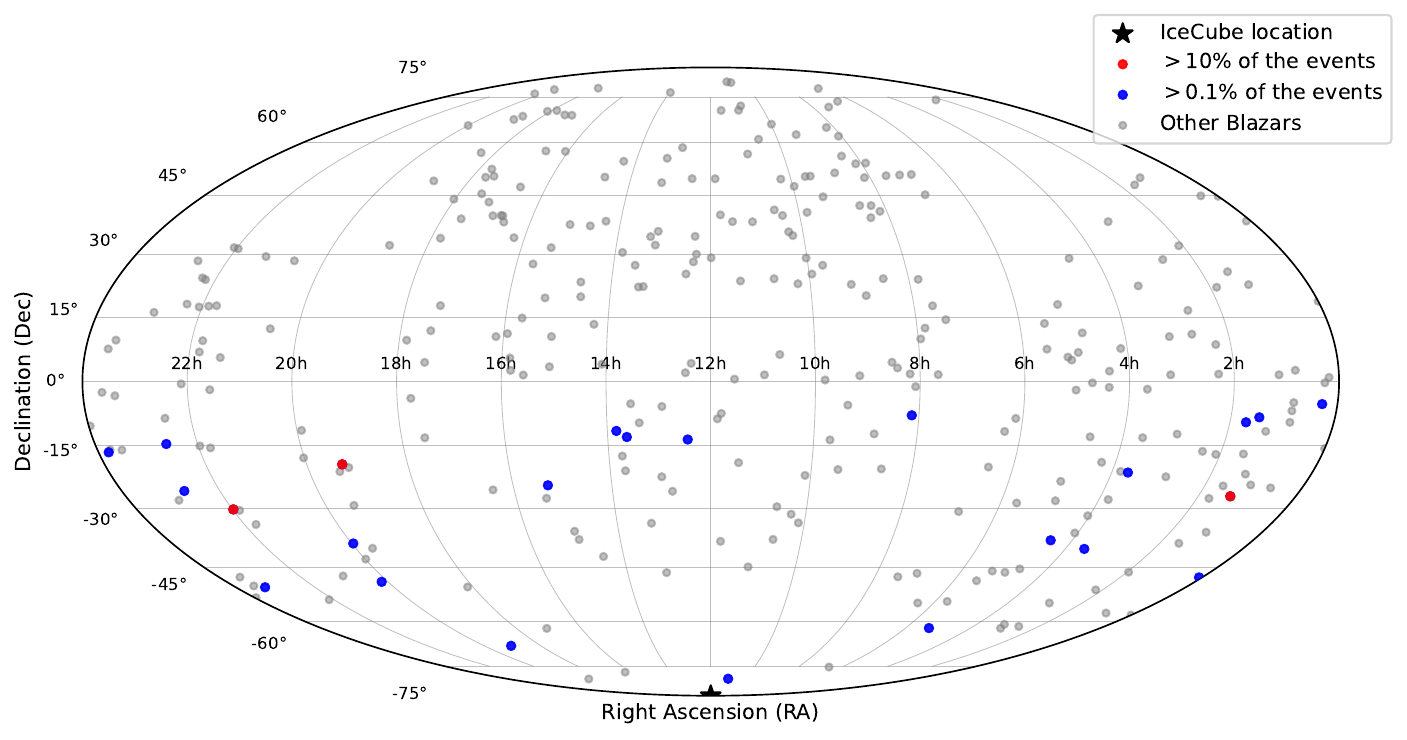}
    \caption{\label{fig:skymap} Sky map with the positions of the sources from~\cite{Rodrigues:2023vbv} that we use in this work. We highlight the blazars that contribute with more than the 10\% (red) and the 0.1\% (blue) of the number of events of BBDM in IceCube, with $\sigma^\mathrm{SI}_{\chi N} = 10^{-30} \, \mathrm{cm}^2$, $m_{Z'} = 10$ MeV and $m_{\chi} = 10$ MeV.}    
\end{figure}
The distribution of blazars considered in this work is shown in \Cref{fig:skymap}. Here, we indicate in red those sources providing more than $10\%$ of the events at \ICns, and in blue those providing at least $0.1\%$ of the events. A first observation is that, as expected, the dominant contribution comes from AGNs in the Southern sky (defined by $\sin(\theta_{DE}) > 0.2$), since the Earth is opaque to BBDM, meaning that the flux coming from Northern sky sources is almost completely absorbed. For $m_{Z'} = 10$ MeV and $m_{\chi} = 10$ MeV, and setting $\sigma^\mathrm{SI}_{\chi N} = 10^{-30} \, \mathrm{cm}^2$, we find that around $\sim 48 \% $ of the events predicted in IceCube are due to the BBDM flux from the blazar 1H~1914-194. This is a consequence of its hard spectrum and its location in the southern hemisphere. The other two blazars most relevant for the number of events are PKS~0118-272 and PKS~2155-304, responsible for $\sim 12.8 \% $ and $\sim 13.1 \% $ of the events, respectively. For other cases, the contributions of each AGN to the total event number in \IC can vary depending on the level of attenuation, determined by the magnitude of the cross section. The case with $\sigma^\mathrm{SI}_{\chi N} = 10^{-30} \, \mathrm{cm}^2$, $m_{Z'} = 10$ MeV and $m_{\chi} = 10$ MeV represents a moderate attenuation case.

\begin{figure}
    \centering
    \includegraphics[width=0.49\linewidth]{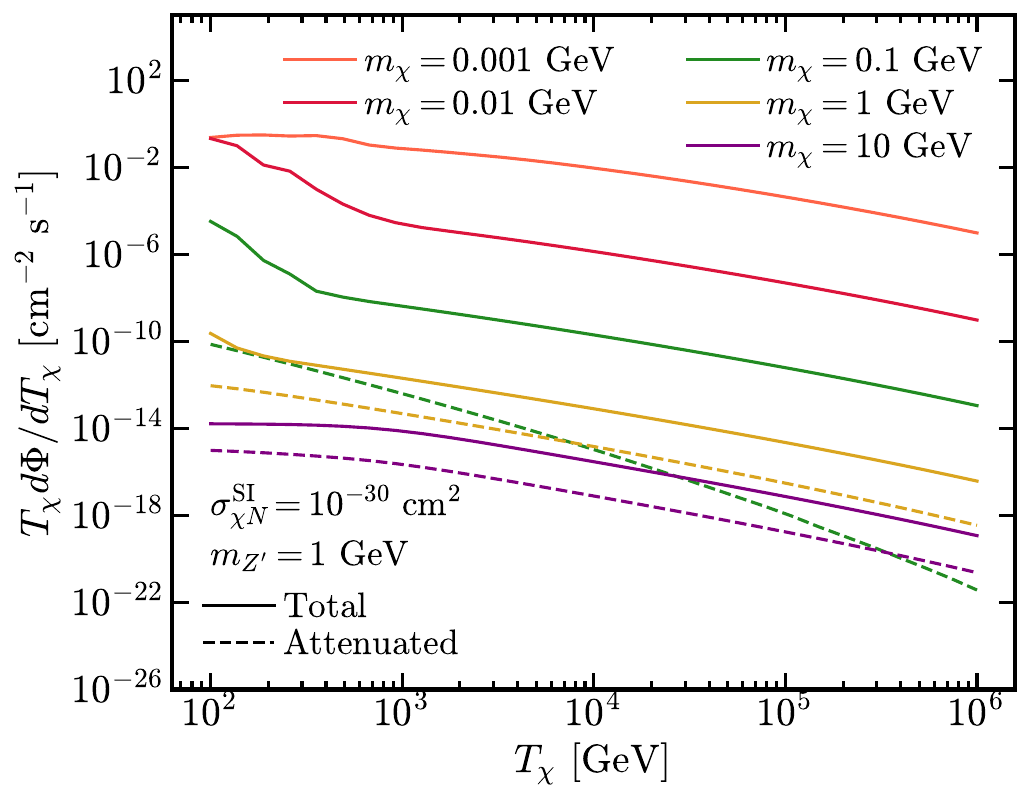}
    \includegraphics[width=0.49\linewidth]{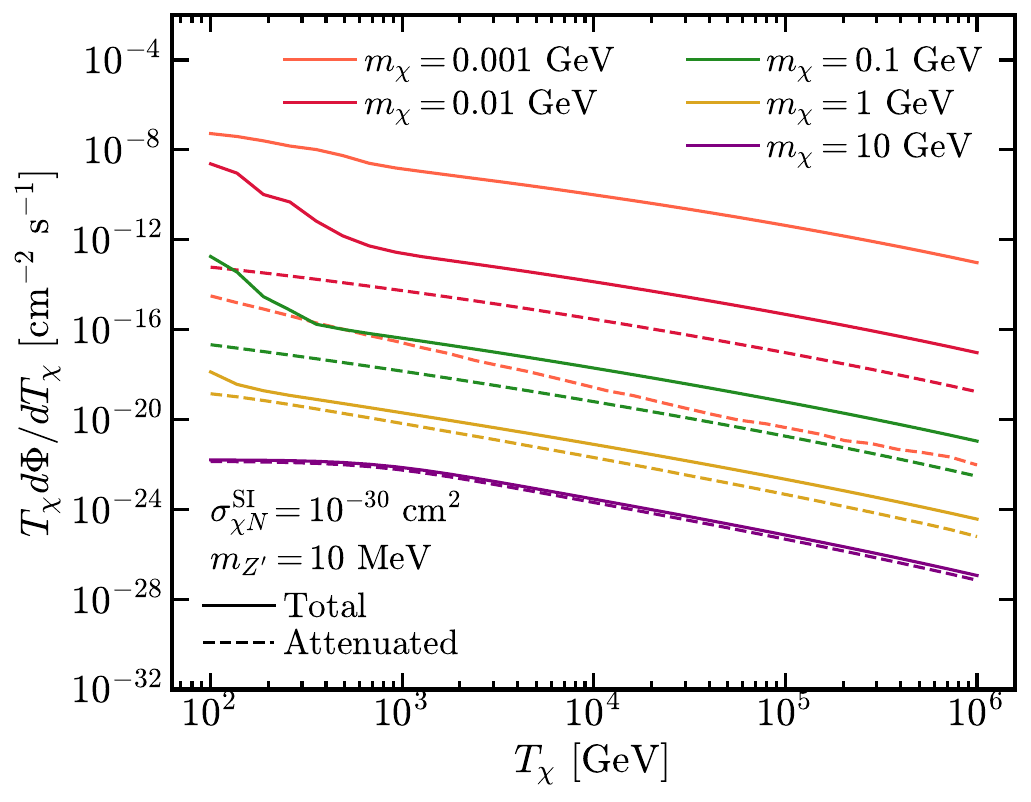}
    \includegraphics[width=0.49\linewidth]{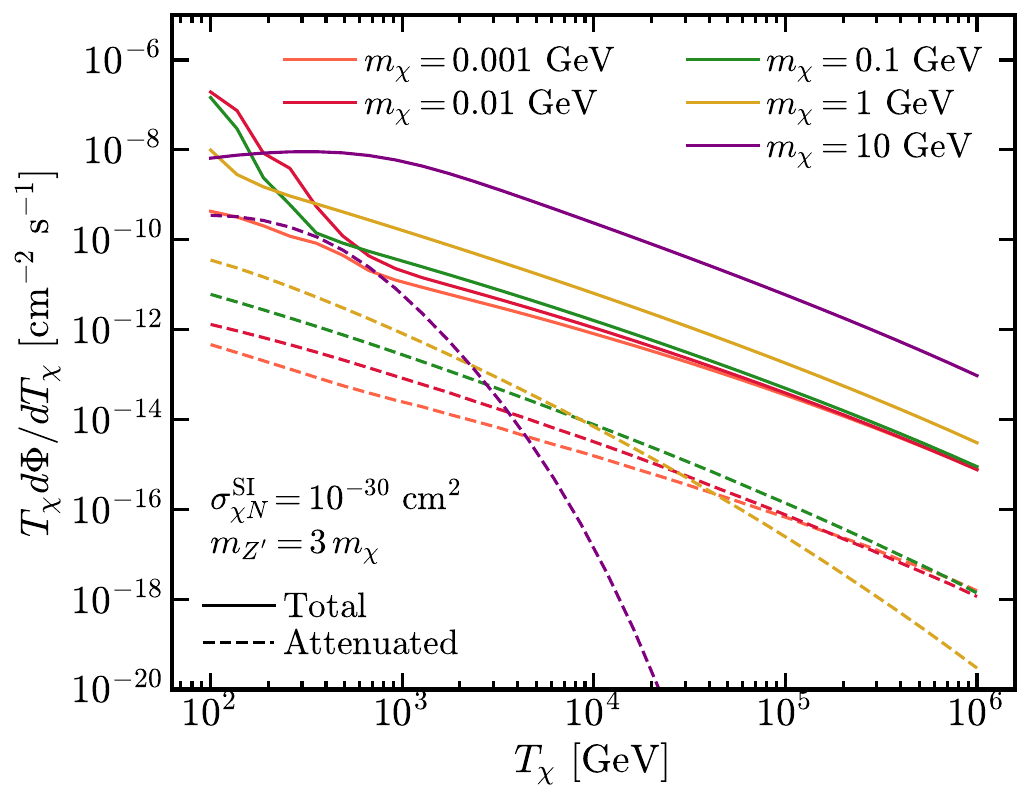}
    \caption{ Fluxes for each of the cases selected for BP1. We show the fluxes as they reach the Earth (solid line), and the attenuated ones (dashed lines), after they cross the Earth to reach the \IC detector. We are fixing the respective non relativistic total cross section to $\sigma^\mathrm{SI}_{\chi N} = 10^{-30} \, \mathrm{cm}^2$. BP2 exhibit the same behavior, but with a rescaling factor of $\mathcal{O} (10^{-1})$. }
    \label{fig:fluxes_NR}
\end{figure}

We show our results in \Cref{fig:fluxes_NR}, which displays the total and attenuated fluxes for the same scenarios shown in \Cref{fig:cs_NR}, for BP1. As done before, we fix the couplings such that the non-relativistic cross section is $10^{-30} \, \mathrm{cm}^2$. The total fluxes (solid lines) represent what arrives to Earth, while the attenuated fluxes (dashed lines) are those arriving to \ICns, after part of the flux is absorbed when traversing the planet.

In general, for a fixed $m_{Z'}$, lighter DM produces higher total fluxes because it is more abundant around the blazar, as the column density is fixed ($n_\chi \equiv \Sigma^s_{\rm DM} / m_\chi$).\footnote{As shown in simulations of DM orbiting around a supermassive black hole, dark matter mass density is a fixed quantity~\cite{Gondolo_1999}.} This is clearly reflected by the solid lines in the upper panels of \Cref{fig:fluxes_NR}. Moreover, in this regime we also find that, for a fixed $m_\chi$, a larger mediator mass consistently leads to a larger total flux, since the propagator in the differential cross section provides a lesser suppression at high values of $Q^2$.

For $m_{Z'}=3m_\chi$ (lower panel), smaller DM masses also lead to smaller mediator masses, so the aforementioned effects compete in opposite directions. We find that, in this regime, in order to have a larger flux it is better to have a larger mediator mass, rather than a small DM mass, particularly at large $T_{\chi}$.

For the fluxes without attenuation, we also find in all panels that RES interactions can dominate over DIS for the lowest values of the kinetic energy, $T_\chi \lesssim 10^3$~GeV, leading to a steep slope. For fixed $m_{Z'}$ we find that, with increasing $m_\chi$, the change in slope happens for smaller values of $T_\chi$. The only exception is found for $m_\chi = 1$~MeV, due to $d\sigma_\mathrm{RES} / dT_\chi$ being kinematically suppressed at these energies. In principle, larger values of $T_\chi$ can avoid this suppression, but at this point we find that DIS is already dominating the total cross section.

Let us now focus on the attenuated fluxes, for fixed mediator masses (upper panels). As shown in \Cref{eq:cascade}, the attenuation depends exponentially on the cross section, which appears in \Cref{fig:cs_NR}. There, we found that lighter DM masses lead to larger cross sections. This implies that, for all values of $E_\chi$ and both values of $m_{Z'}$, DM with a mass of 1 MeV would be more attenuated than one with a mass of 10~GeV by a factor $\sim 10^6$ in the exponent. Similarly, if $m_\chi=1$~GeV, the attenuation would lead to a factor $\sim 10$ on the exponent with respect to DM with $m_\chi=10$~GeV. Thus, light DM is severely more suppressed than heavier DM, unless the nucleon column density $\tau$ vanishes.

Indeed, on the upper left panel of \Cref{fig:cs_NR}, where $m_{Z'}=1$~GeV, we find that the fluxes for $m_\chi=1,\,10$~MeV are so attenuated that they effectively vanish. One should not expect a uniform suppression, though, as different sources with different attenuation contribute differently to the spectrum. For $m_{Z'}=10$~MeV (upper right panel), the cross sections are not as large, but even then we find a very strong suppression of the flux for light DM.

For our third scenario, with $m_{Z'}=3m_\chi$, the cross sections shown in \Cref{fig:cs_NR} increase with $m_\chi$ at large energies. Thus, on the lower panel of \Cref{fig:fluxes_NR}, we find that the high energy flux is severely suppressed for large DM masses. For low energy the cross sections are of the same order of magnitude, so the attenuation on this end of the spectrum is similar in all cases, for example, at $E_\chi=1$~TeV all fluxes are suppressed by around two orders of magnitude.

The RES contribution to the cross section that generates the low-energy structures in the unattenuated flux also increases the interaction probability during propagation through the Earth. Consequently, these structures are largely removed from the attenuated flux before reaching \ICns.

One can thus conclude that cross sections have a dual role in the calculation of the fluxes. On the one hand, the production of the boosted DM beam is proportional to the differential cross section, $d\sigma / dT_\chi \propto d\sigma / dQ^2$, meaning that a larger cross section will lead to a larger total flux. However, too large couplings also impact the attenuation, meaning that very small and very large couplings may converge to the same result: vanishing fluxes at \ICns.

These behaviors are roughly the same for both benchmarks, BP1 and BP2, up to a constant factor that rescales the fluxes.


\section{Constraints from IceCube}
\label{sec:4_IC}

We search signals of BBDM at \IC using $641$ days of data reported in~\cite{IceCube:2014rwe}. This reference reports measurements of astrophysical and atmospheric neutrino fluxes down to energies of approximately 1~TeV. This dataset was chosen because of the relatively low energies considered since, as we see on \Cref{fig:fluxes_NR}, the total DM fluxes reaching the \IC detector fall with energy.

With the total DM flux reaching the \IC detector, we can estimate the number of events observed due to BBDM.
As the DM signals are almost indistinguishable from neutral current neutrino events, we
follow~\cite{Guo:2020drq} and calculate the effective area for DM detection $A_{{\rm eff},\chi}$ as a function of incoming DM energy $E_\chi$, declination angle of the AGN $\theta_{DE}$ and deposited energy $E_{\rm dep}$  using
\begin{eqnarray}
A_{{\rm eff},\chi} (E_\chi,\,\theta_{DE},\,E_{\rm dep}) = A^{NC}_{{\rm eff},\nu_e} (E_\chi,\,\theta_{DE},\,E_{\rm dep}) \frac{ \sigma_{\chi p}(E_\chi)}{ \sigma_{\nu_e p}^{NC}(E_\chi)}
\end{eqnarray}
where $A^{NC}_{{\rm eff},\nu_e}$ is the effective area for electron neutrinos interacting via neutral currents and producing a cascade-like interaction, and $\sigma_{\chi p}$ ($\sigma_{\nu_e p}^{NC}$) is the total DM - proton (neutral current neutrino - proton) cross-section. For $A^{NC}_{{\rm eff},\nu_e}$ and $\sigma_{\nu_e p}^{NC}$ we use data published by \IC in~\cite{IceCube:2014rwe}, and the neutrino cross sections published in~\cite{Connolly:2011vc}, respectively.

With this information, the number of events observed in \IC from a single AGN on a bin around $\theta_{DE}$, as function of $E_{\rm dep}$, can be calculated. Given that the effective area in~\cite{IceCube:2014rwe} is given in terms of bins of size $\Delta E_\chi= 0.07\,E_{\chi} \, \log(10)$, with $E_{\chi}$ being the central value of the energy bin, and $\Delta s_{DE}=0.2$, with $s_{DE}$ being the sine of the declination angle, the number of events comes from:
\begin{equation}
\frac{dN}{dE_{\rm dep}}(E_{\rm dep},\,\theta_{DE}) = 2\pi \,t_{\rm exp} \sum_{E_\chi} \, A_{{\rm eff},\chi} (E_\chi,\,\theta_{DE},\,E_{\rm dep}) \, \phi^{IC}_{\chi}(E_\chi,\,\theta_{DE}) \, \Delta E_\chi \, \Delta s_{DE}
\end{equation}
where $t_{\rm exp}=641$~days is the exposition time. Finally, the total number of events is the sum of the contributions of each AGN.

\begin{figure}[tp]
    \centering
    \includegraphics[width=0.99\linewidth]{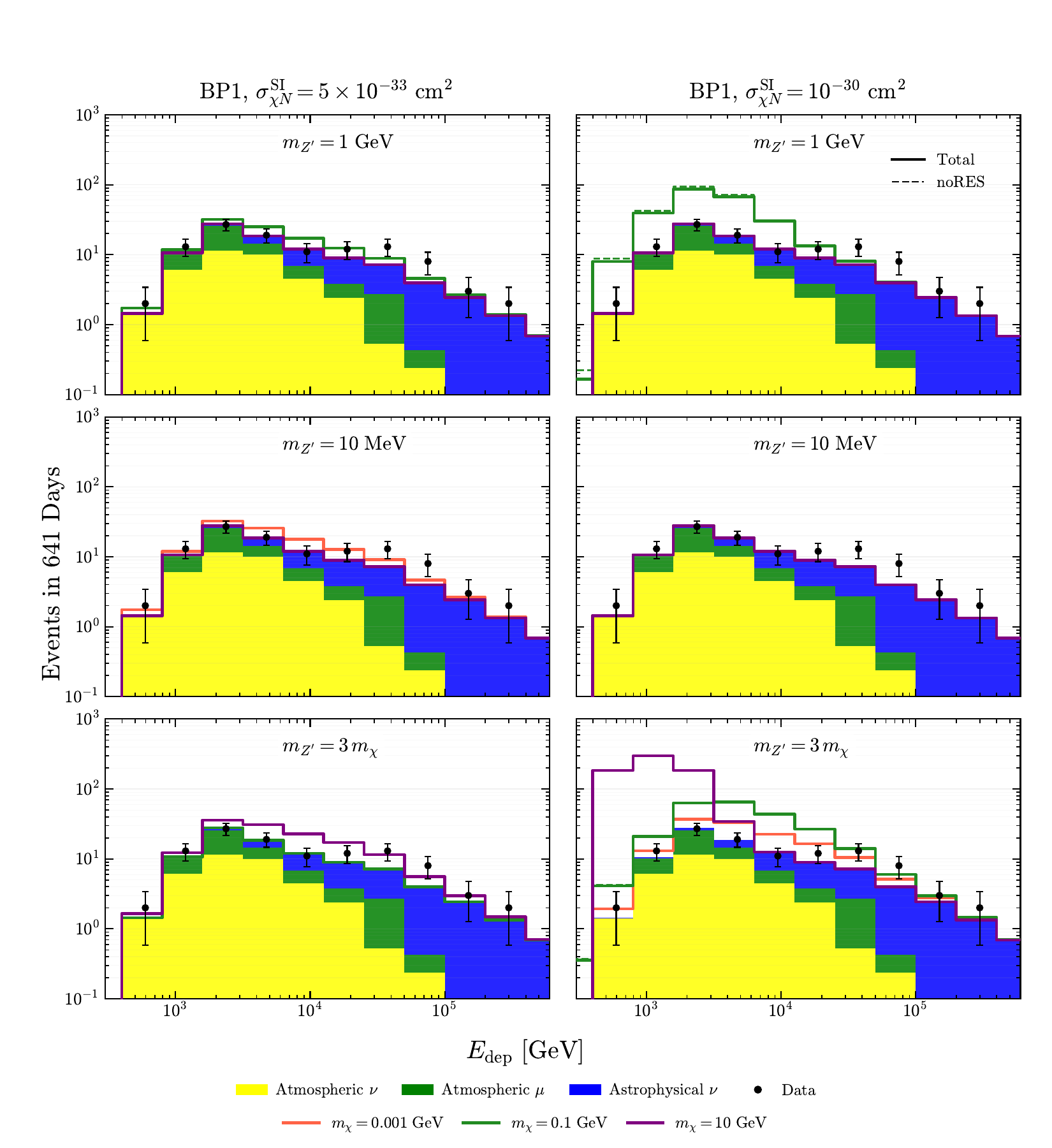} \\
    \caption{\label{fig:histos_NR1e-30_BP1}
    Spectrum of events at \ICns, taking $\sigma^\mathrm{SI}_{\chi N} = 5\times10^{-33}~{\rm cm^2}$ ($10^{-30}~{\rm cm^2}$) on the left (right) panels. The original fluxes correspond to those from BP1. Results with $\sigma^\mathrm{SI}_{\chi N} = 10^{-30} {\rm cm^2}$ and BP1, with all the contributions in solid lines and only DIS in dashed lines, in the production, attenuation and detection of the BBDM flux.}
\end{figure}

In \Cref{fig:histos_NR1e-30_BP1} we show the predicted number of events as a function of $E_{\rm dep}$. We show predictions for several values of $m_\chi$, setting $\sigma^\mathrm{SI}_{\chi N}=5\times10^{-33},\,1\times10^{-30}$~cm$^2$ (left, right panels), with our usual assumptions for mediator masses on the upper, center and lower panels. These are compared against the fitted background and the observed events, as reported in~\cite{IceCube:2014rwe}. The background is comprised by atmospheric neutrinos, atmospheric muons, and astrophysical neutrinos, which on all panels are shown in yellow, green and blue, respectively.

For $m_{Z'}=1$~GeV (upper panels of \Cref{fig:histos_NR1e-30_BP1}) we find that the signal events exceed the background only when $m_\chi\approx0.1$~GeV. This is in agreement with \Cref{fig:fluxes_NR}, where we showed that even though very light DM masses had larger total fluxes, the attenuation by the Earth brought those fluxes to negligible values. For larger masses the attenuation effects were small, but then we were faced by a smaller total flux, and by smaller cross sections at the time of detection.

For $m_{Z'}=10$~MeV (center panels of \Cref{fig:histos_NR1e-30_BP1}), only scenarios with very light DM masses can produce a signal over the background and, perhaps counterintuitively, only if the couplings are small enough. As seen in \Cref{fig:fluxes_NR}, even though the total fluxes for this mediator are much lower than the previous case, the attenuation is also milder, meaning that fluxes for masses smaller than $0.1$~GeV do not vanish. Since these also have larger cross sections, they are the only situations where DM has a chance of being observed. Still, if the couplings are too large (i.e.\ $\sigma^\mathrm{SI}_{\chi N}\gtrsim10^{-30}$~cm$^2$), the attenuation becomes too strong again, and the signal is lost.

Finally, for $m_{Z'}=3m_\chi$, we remind the reader that in this scenario the total and attenuated fluxes for $m_\chi\lesssim1$~GeV are relatively similar. This was due to all cases having relatively similar cross sections, where the enhancement coming from small $m_\chi$ was balanced by a suppression due to a similarly small $m_{Z'}$. Thus, our results (lower panels of \Cref{fig:histos_NR1e-30_BP1}) find that all DM masses predict a large number of signal events, as long as $\sigma^\mathrm{SI}_{\chi N}\sim10^{-30}$~cm$^2$. For smaller cross sections, the signal events vanish, with the exception of those for $m_\chi\sim10$~GeV, where the cross section is enhanced (see \Cref{fig:cs_NR}).

An additional feature worth mentioning in \Cref{fig:histos_NR1e-30_BP1} is that, whenever applicable, we show the event spectrum calculated without RES scattering, that is, by turning off all RES interactions in production, attenuation and detection. We see that the inclusion of RES does not usually have any effect on the number of events, which is understood from our discussion around \Cref{fig:fluxes_NR}, that is, all enhancements at production are lost due to attenuation. The exception happens for $m_{Z'}=1$~GeV and $m_\chi=0.1$~GeV, where we find that turning on RES leads to a slight reduction of the number of events of around $\sim 9 \%$. So, at the end of the day, the additional losses from attenuation are stronger than the gains at production and detection.
In the rest of the cases, the impact on the number of events due to the inclusion of RES is below $\sim 3 \%$.

Having calculated our expected signal due to BBDM, we can compare our results with the events observed by \ICns, reported in~\cite{IceCube:2014rwe}. 
The reported data is divided between the northern and southern sky, with the southern sky defined by $\sin(\theta_{\rm DE}) > 0.2$. As mentioned before, we consider only southern sky events, as for the relevant cross sections analyzed the BBDM coming from the northern sky is almost completely attenuated before reaching the detector.
With this, for each value of $m_\chi$ and $m_{Z'}$, we use the CLs method, briefly outlined in \Cref{sec:CLsMethod}, to place bounds on $\sigma^{\rm SI}_{\chi N}$ (see \Cref{sssec:non_rel_cs}).

\begin{figure}
    \centering
    \includegraphics[width=0.49\linewidth]{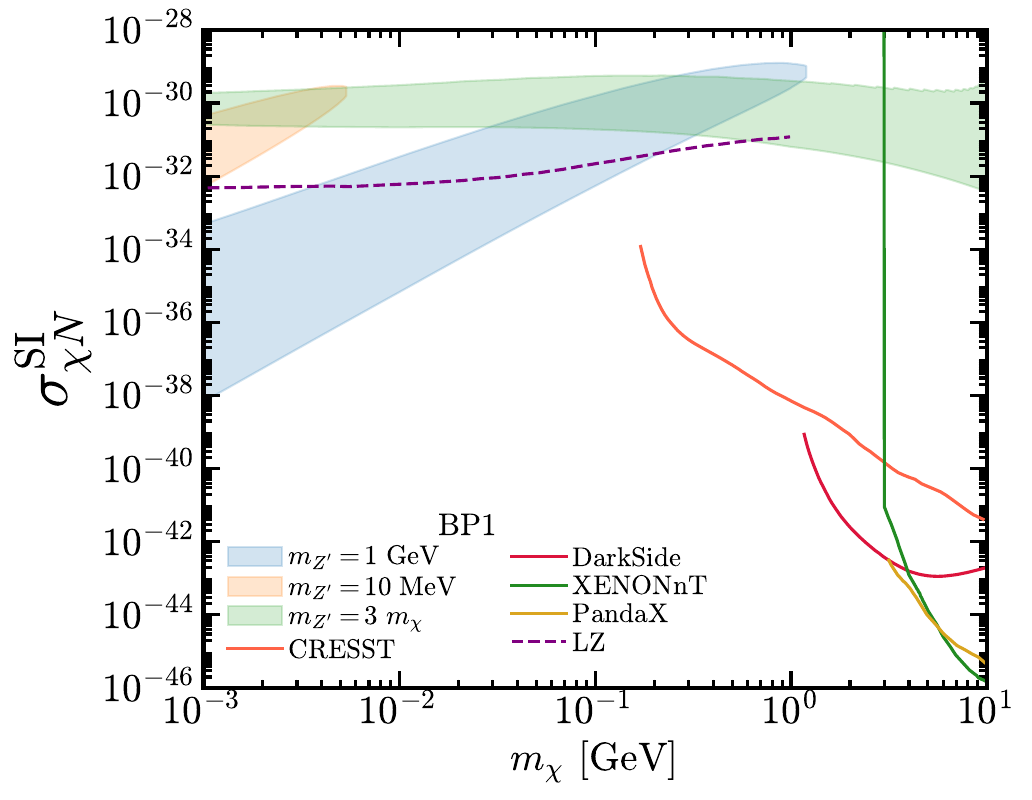}
    \includegraphics[width=0.49\linewidth]{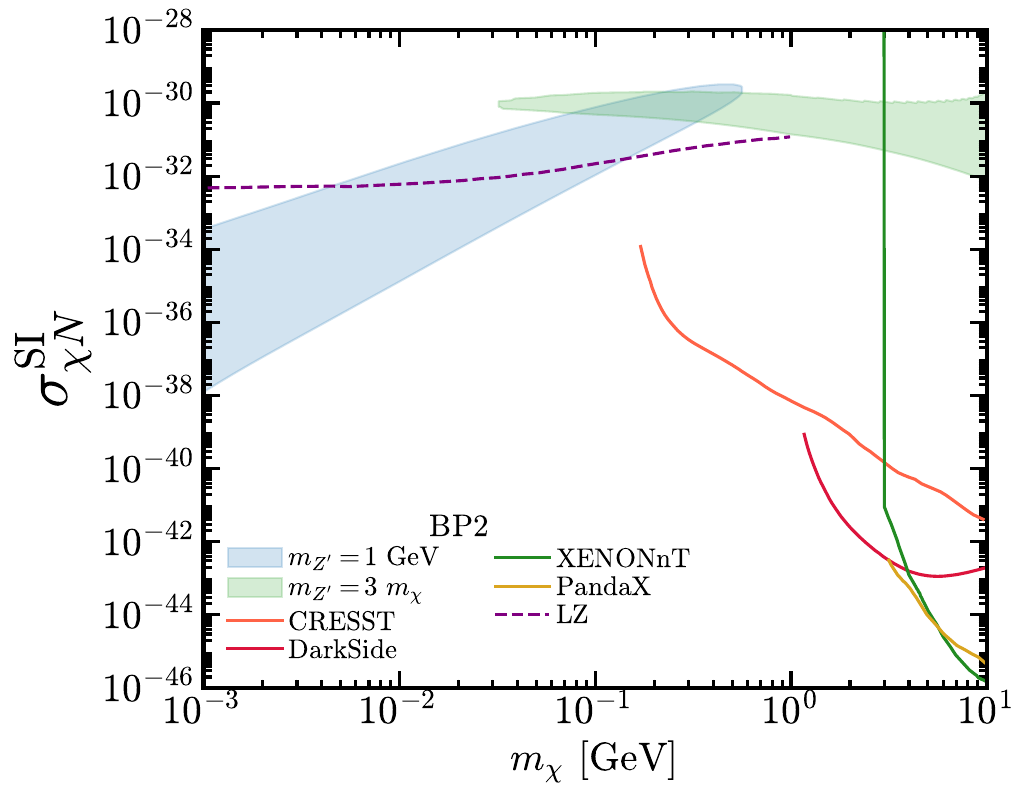}
    \caption{\label{fig:plots_uplim_zexcl_vect_cls} Exclusion regions for the cases presented in this work, $\sigma^{\rm SI}_{\chi N}$, for both DM spike benchmarks, compared with the spin independent direct detection limits. On BP2, we cannot place any bounds when $m_{Z'}=10$~MeV.}
\end{figure}

We show our final results as exclusion regions on \Cref{fig:plots_uplim_zexcl_vect_cls}, for BP1 and BP2 on the left and right panels, respectively. These are compared with recent direct detection DM bounds on spin independent cross sections~\cite{DEAP:2019yzn, CRESST:2019jnq, PandaX:2022aac, PandaX:2022xqx, DarkSide-50:2022qzh, PandaX:2024qfu, LZ:2024zvo, XENON:2025vwd}\footnote{Data extracted from \url{https://github.com/cajohare/DirectDetectionPlots}}, as well as the latest limits from LZ~\cite{LZ:2025iaw}.

We observe that the exclusion regions exhibit a band-like structure, which is again consistent with the results shown in previous Figures. In general, the lower limit in the band arises from a too low flux being originally produced, indeed, if the coupling is too small it is unlikely for the protons in the blazar to boost a considerable amount of DM. The upper limit, on the other hand, is attributed to a too large attenuation, as presented in \Cref{fig:fluxes_NR,fig:histos_NR1e-30_BP1}.

For fixed mediator masses, $m_{Z'}=10$~MeV and 1~GeV, shown respectively in red and blue in \Cref{fig:plots_uplim_zexcl_vect_cls}, we find that the sensitivity increases for lower DM masses. As explained previously, lower $m_\chi$ leads to both larger incoming fluxes but a larger attenuation. The larger flux helps bring the lower bound to smaller values of $\sigma^\mathrm{SI}_{\chi N}$, but the larger attenuation also decreases its upper bound. Fortunately, the gain in the larger flux seems to outweigh the loss due to stronger attenuation, leading to the bands having a larger width at small masses.

Another feature of these scenarios is that heavier mediator masses are more strongly constrained, since this produces higher cross sections at higher energies. Therefore, the scenario where $m_{Z'}=1$~GeV imposes more stringent constraints than that for $m_{Z'}=10$~MeV across the evaluated parameter space.

The third scenario, $m_{Z'}=3m_\chi$, shown in green in the Figure, is particularly interesting, as the mediator mass is always larger than the DM mass. This leads to the bands having a completely different behavior. As demonstrated in Figure~\ref{fig:histos_NR1e-30_BP1}, the signal is strongest for large $m_\chi$, but it is non-negligible for smaller DM masses too. In BP1, this leads to a nearly horizontal band stretching across our parameter space, broadening at large masses due to a larger cross section at the moment of detection.

When comparing BP1 and BP2, we find that the latter benchmark has weaker limits. Here, we find that the $m_{Z'}=10$~MeV case cannot probe the evaluated parameter space at all, while for $m_{Z'}=1$~GeV the maximum DM mass that can be constrained is $\sim0.5$~GeV. Moreover, for $m_{Z'}=3m_\chi$ the horizontal band is cut off for masses lower than $\sim30$~MeV. The reason for this is that, since $R_{\rm BLR}\gg100 R_s$, BP2 has a much smaller DM column density than BP1, meaning there exist less DM particles that can be boosted from by the proton flux. This in turn leads to lower fluxes.

Comparison with bounds from direct detection experiments confirm the complementarity of searches performed by \ICns, which indirectly probe masses smaller by many orders of magnitude. It is worth noting that even though the LZ bound, shown dashed, seems to already rule out our exclusions at $m_{Z'}=10$~MeV and $m_{Z'}=3m_\chi$, the former limit has been generated with a constant spin-independent non-relativistic cross section, multiplied by a nucleus form factor, meaning that its interpretation is not straightforward~\cite{Yu:2025bwu}.



\section{Conclusions}
\label{sec:4_conclusions}

In the present work, we have investigated the direct detection of boosted fermionic DM interacting with quarks through a massive vector mediator, $Z'$. Motivated by the possibility that astrophysical environments can generate highly energetic DM fluxes, we considered the acceleration of these particles by proton beams in active galactic nuclei. In particular, adopting the Gondolo--Silk profile for the dark matter distribution around supermassive black holes, we estimated the boosted DM fluxes produced from a sample of 324 blazars. We then evaluated the corresponding fluxes expected at \ICns, consistently accounting for terrestrial attenuation effects as well as the detector response. By comparing the resulting event predictions with the observed high-energy neutrino flux at \ICns, we derived exclusion limits on the parameter space of this scenario.

Our analysis includes the following features: (1) we include elastic, resonant and deep inelastic interactions, (2) we compare two conservative assumptions regarding the astrophysical region where DM is accelerated (starting at $100R_S$ or $R_{\rm BLR}$), (3) we build the DM flux from a sample of 324 blazars taken from a catalog, and (4) given the mainly inelastic nature of our interactions, we applied the cascade equation in order to take into account attenuation effects.

In agreement with the literature, our results show that the detection of these scenarios at \IC can become significantly more competitive than conventional direct-detection experiments in the low-mass regime. As is well known, the origin of this enhanced sensitivity can be understood from the limitations affecting standard nuclear recoil searches, which rapidly lose efficiency for DM masses below $\mathcal{O}(1)$~GeV due to the small recoil energies involved. In contrast, the boosted DM flux considered here scales inversely with its mass, as lighter DM particles correspond to a larger number density in the environments surrounding blazars, increasing the available population that can be accelerated by the proton jets. In our best case scenarios, we can reach sensitivities of $\sigma^\mathrm{SI}_{\chi N} \sim 10^{-38} \, \mathrm{cm}^2$ for DM masses of 1~MeV and $\sigma^\mathrm{SI}_{\chi N} \sim 10^{-32} \, \mathrm{cm}^2$ for 1~GeV. 

We have also found that the dominant contribution to the signal originates from deep inelastic scattering processes. As a consequence, the analysis exhibits greater sensitivity to comparatively heavier mediator masses, as expected. This behavior can be traced back to the momentum-transfer dependence introduced by the mediator propagator. For lighter mediators, the cross section becomes strongly suppressed at large values of the momentum transfer $Q^2$, reducing the contribution from high-energy scattering events. Since deep inelastic scattering processes become more efficient when larger momentum transfers are kinematically accessible, heavier mediator masses lead to less suppression and therefore to larger detectable event rates.

In addition, we have shown that resonance scattering processes, in which a single pion is produced in the final state, can be locally relevant in the lower-energy region accessible to \ICns. Although deep inelastic scattering dominates the total event rate at higher energies, resonance production can provide a contribution close to the experimental threshold by enhancing the fluxes for those energies and by raising the cross sections for detection. However, resonant scattering also affects attenuation, thus decreasing the overall effect to almost negligible levels. The impact on the final event rate is small for the model and masses under consideration: below approximately 3$\%$ in most of the benchmarks studied, with a maximum reduction of about 9$\%$. For other masses and for different experimental set-ups, where attenuation may have a weaker effect, this consequence is not necessarily true.

Finally, we find that, out of the 324 sources, due to attenuation effects only three are responsible for about $70\%$ of the signal events. 

\acknowledgments
J.~H.~Z. is supported by the National Science Centre, Poland (research grant No. 2021/42/E/ST2/00031).
G.~D.~Z. is supported by Universidad Nacional Autónoma de México Postdoctoral Program (POSDOC).
G.~D.~Z. also acknowledges funding from the Vicerrectorado de Investigación at Pontificia Universidad Católica del Perú via the Estancias Posdoctorales en la PUCP 2023 program.
A.~G.~M. and J.~J.~P. acknowledge funding by the \textit{Dirección de Gestión de
la Investigación} at PUCP, through grant DFI-PUCP-PI1144. 


\appendix

\section{Non-relativistic elastic interactions}
\label{sssec:non_rel_cs}

For comparison purposes, it is useful to compute the cross sections that are relevant for standard DM direct detection experiments at very low energies. In this regime, we can rewrite the interaction by extracting the propagator of the mediator as $\sim 1 / m_{Z^\prime}^2$, 
thus allowing the amplitude to be obtained from an effective interaction term:
\begin{equation}
\begin{split}
\mathcal{L}_\mathrm{eff} \supset \frac{g_D g_{NZ^\prime}}{m_{Z^\prime}^2}  \sum_q c_V^q \left[ \overline{\chi} \gamma^\mu \chi \right] \left[\overline{q} \gamma_\mu q \right]\,.
\end{split}
\end{equation}
As a first step, let us define $\mathcal{O}^q_V \equiv c^q_V (\overline{\chi} \gamma^\mu \chi)(\overline{q} \gamma_\mu q)$ as our purely vector operator. As worked out in~\cite{Fitzpatrick:2012ix, Cirelli:2013ufw, Catena:2015uha}, we can connect these effective operators at the quark level to ones involving nucleons. The operator is $\mathcal{O}^N_V \equiv c^N_V (\overline{\chi} \gamma^\mu \chi)(\overline{N} \gamma_\mu N)$, where $c^p_V = 2 c^u_V + c^d_V$ and $c^n_V = c^u_V + 2c^d_V$.

Finally, we can transform these relativistic operators into fully non-relativistic ones:
\begin{align}
    \mathcal{O}^N_V \to \mathcal{O}^\mathrm{NR}_V \equiv 4 \, c^N_V m_\chi m_N\,.
\end{align}
Since there is no dependence on spin, the former is usually known as producing a spin-independent cross section. Our non-relativistic operator, $\mathcal{O}^\mathrm{NR}_V$, yields the same cross sections for any nucleon under the approximation of equal nucleon masses ($m_p \sim m_n$). Therefore, we will speak of a spin-independent cross section with nucleons ($\sigma^\mathrm{SI}_{\chi N}$).

As follows from \cite{Lebedev:2014bba, Borschensky:2020olr}, the final result for these cross section is:
\begin{equation}
\label{eq:NR_CrossSection}
\begin{split}
\sigma^\mathrm{SI}_{\chi N} &= \frac{\mu^2_{\chi N}}{\pi} \left( \frac{g_D\, g_{NZ^\prime}}{m_{Z^\prime}^2} \right)^2 \left(c_V^\chi \right)^2 \left(c_V^N \right)^2\,,
\end{split}
\end{equation}
where $\mu_{\chi N} \equiv m_\chi m_N / (m_\chi + m_N)$ is the reduced mass of the system DM - nucleon.
This cross section will be used to explore the parameter space of direct detection at high energies and put them in the context of low energetic bounds.

\section{Kinematic relations between cross section parameters and scattering angle}
\label{app:kinematic_rels}

\begin{figure}[tbp]
    \centering
    \includegraphics[width=0.49\linewidth]{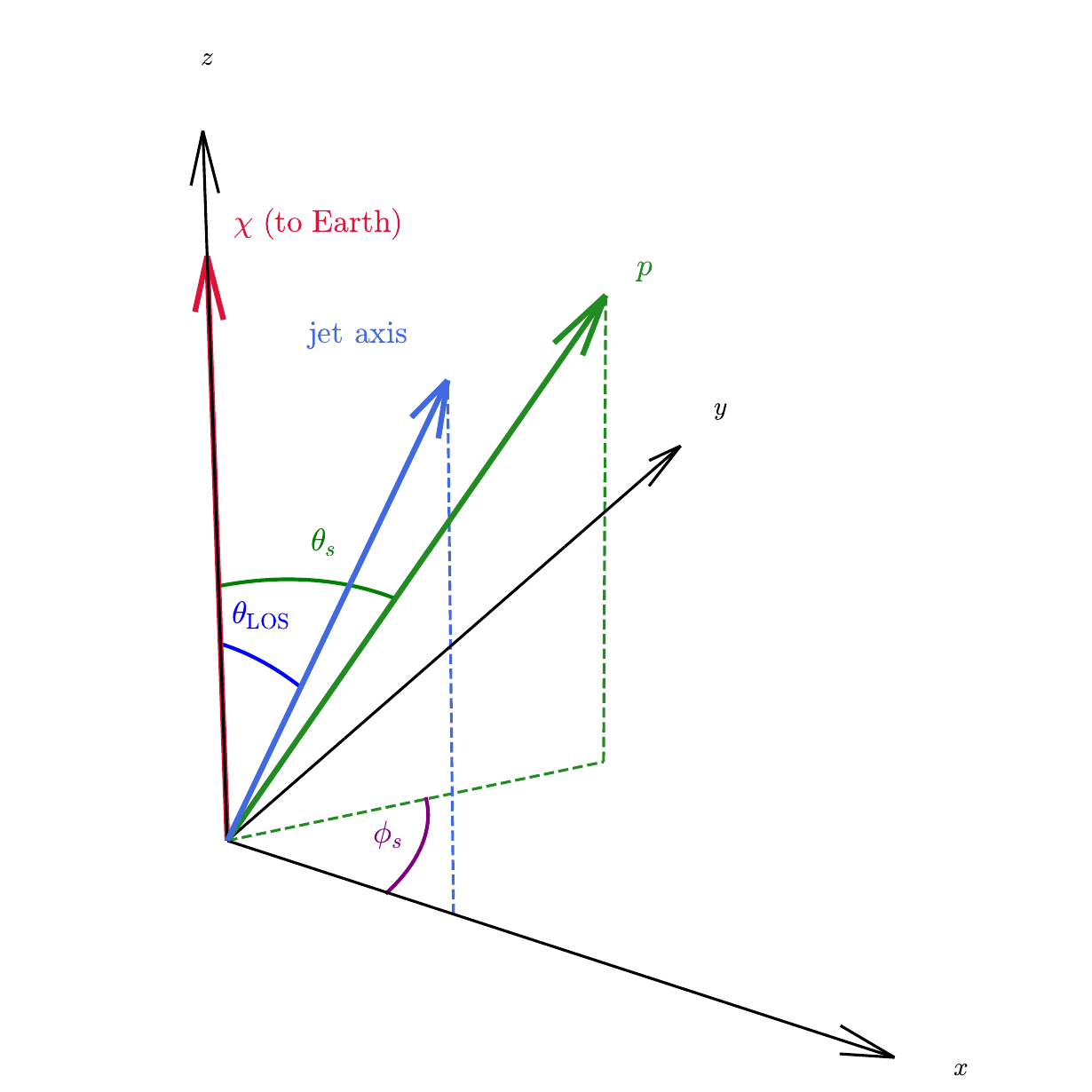}
    \caption{\label{fig:geometry_interaction} Vectors defining the interaction where protons $p$ boost dark matter particles $\chi$. The outgoing $\chi$ points towards the Earth and defines the $Z$-axis, while the central axis of the blazar jet lies on the $XZ$ plane, making an angle $\theta_\mathrm{LOS}$ with the outgoing $\chi$. The proton within the jet has a solid angle defined by $\theta_s$ and $\phi_s$ with respect to $\chi$. This proton is also at an angle $\theta$, not shown, with respect to the central axis of the jet.} 
\end{figure}
In order to compute the flux of boosted DM, we need to compute $\mu \equiv \cos \theta$, where $\theta$ is the angle between the incoming proton and the direction of motion of the blob frame, with respect to observers on Earth. This quantity also depends on the scattering angles defining $d\Omega$, $\mu_s$ and $\phi_s$, and the line-of-sight, $\theta_\mathrm{LOS}$, as can be seen in Eq. (22) in \cite{Wang:2021jic} and is also shown in \Cref{fig:geometry_interaction},
\begin{equation}
    \mu \left(\mu_s, \phi_s, \theta_\mathrm{LOS} \right) = \mu_s \cos \theta_\mathrm{LOS} + \sin \phi_s \sin \theta_\mathrm{LOS} \sqrt{1 - \mu_s^2}\,.
\end{equation}

The key quantity to be found is $\mu_s$, which depends on the kinematic variables used to compute the cross section of each process. Therefore, we will have $\mu^k  \equiv \mu \left[\mu_s^{k} \left( \{\Pi_k\} \right), \phi_s, \theta_\mathrm{LOS} \right]$, where $k =$ EL, RES, DIS, and where $\Pi_k$ are the kinematic variables used to parametrize the cross section $k$. The precise relations are,
\begin{equation}
\begin{split}
\mu_s^\mathrm{EL} \left( T_p, T_\chi \right) &= \frac{ \left(m_p + m_\chi + T_p \right) T_\chi}{\sqrt{T_p T_\chi \left( 2m_p + T_p \right) \left( 2m_\chi + T_\chi \right)}}\,,\\
\mu_s^\mathrm{RES} \left( T_p, T_\chi, W \right) &= \frac{ 2 \left(m_p + m_\chi + T_p \right) T_\chi + W^2 - m_p^2}{2\sqrt{T_p T_\chi \left( 2m_p + T_p \right) \left( 2m_\chi + T_\chi \right)}}\,,\\
\mu_s^\mathrm{DIS} \left( T_p, T_\chi, \nu \right) &= \frac{m_p T_\chi + T_p T_\chi + m_p \nu}{\sqrt{T_p T_\chi \left( 2m_p + T_p \right) \left( 2m_\chi + T_\chi \right)}}\,.
\end{split}
\end{equation}

These functions are used to compute the proton flux when the BBDM fluxes are calculated.

\section{CLs method analysis } 
\label{sec:CLsMethod}

Our exclusion curves are based on the CL$_s$ method~\cite{Read:2000ru, Read:2002hq,Cowan:2010js}. We model the expected number of events in bin $i$ as
\begin{equation}
  \lambda_i(\mu,\vec{\theta})
  = \mu\,s_i
  + (1+\theta_\nu)\,b_i^{\nu}
  + (1+\theta_\mu)\,b_i^{\mu}
  + (1+\theta_a)\,b_i^{a},
\end{equation}
where $s_i$ is the predicted DM signal for unit signal strength $\mu$,
and $b_i^{\nu}$, $b_i^{\mu}$, $b_i^{a}$ are the expected backgrounds
from conventional atmospheric neutrinos, atmospheric muons, and
astrophysical neutrinos, respectively, taken from
Ref.~\cite{IceCube:2014stg}.
The parameters $\vec{\theta} = (\theta_\nu, \theta_\mu, \theta_a)$
are nuisance parameters encoding the normalization uncertainties of each
background component.
The full likelihood is
\begin{equation}
  \mathcal{L}(\mu,\vec{\theta}) \;=\;
  \prod_{i=1}^{20}
  \frac{\lambda_i(\mu,\vec{\theta})^{n_i}\, e^{-\lambda_i(\mu,\vec{\theta})}}{n_i!}
  \;\cdot\;
  \prod_{k \in \{\nu,\mu,a\}}
  \exp\!\left(-\frac{\theta_k^2}{2\sigma_k^2}\right),
  \label{eq:likelihood}
\end{equation}
where $n_i$ are the observed counts, and the second factor is a Gaussian prior for the nuisance parameters. 
The  uncertainties are taken from the best-fit values
reported in Ref.~\cite{IceCube:2014stg}:
$\sigma_\nu = 10\%$ for conventional atmospheric neutrinos,
$\sigma_\mu = 23\%$ for atmospheric muons, and
$\sigma_a   = 17\%$ for astrophysical neutrinos.

The test statistic is the profile likelihood ratio~\cite{Cowan:2010js}
\begin{equation}
  \tilde{q}_\mu =
  \begin{cases}
    -2\ln\dfrac{
      \mathcal{L}(\mu,\,\hat{\hat{\vec{\theta}}}(\mu))}{
      \mathcal{L}(\hat\mu,\,\hat{\vec{\theta}})}
    & \hat\mu \leq \mu, \\[8pt]
    0 & \hat\mu > \mu,
  \end{cases}
\end{equation}
where $(\hat\mu,\hat{\vec{\theta}})$ maximize the likelihood globally
and $\hat{\hat{\vec{\theta}}}(\mu)$ maximize it at fixed $\mu$.
Upper limits at 95\% CL are obtained from the CL$_s$ criterion
\begin{equation}
  \mathrm{CL}_s(\mu) \equiv
  \frac{p_{s+b}(\mu)}{1-p_b} \leq 0.05,
\end{equation}
where $p_{s+b}$ ($p_b$) is the $p$-value of $\tilde{q}_\mu^{\,\rm obs}$
under the signal-plus-background (background-only) hypothesis.
Dividing by $1-p_b$ prevents spurious exclusion of signals to which the
experiment has no sensitivity~\cite{Read:2002hq}.
The signal strength limit $\mu_{\rm lim}$ is obtained by scanning $\mu$
and refining the crossing point CL$_s = 0.05$ with a bisection algorithm.
The statistical model is implemented using
\texttt{pyhf}~\cite{Heinrich:2021gyp}.
A given point in the model parameter space is considered excluded at
95\% CL if $\mu_{\rm lim} < 1$, i.e.\ if the predicted signal strength
exceeds what the data allow.

\bibliographystyle{apsrev4-1}
\bibliography{biblio}{}
\end{document}